\documentclass[traditabstract]{aa} % 
\usepackage{txfonts}
\usepackage{graphicx}
\usepackage{natbib}
\usepackage{color}

\newcommand{\hwvir}{\mbox{\object{HW\,Vir}}}
\newcommand{\huaqr}{\mbox{\object{HU\,Aqr}}}
\newcommand{\nnser}{\mbox{\object{NN\,Ser}}}
\newcommand{\qsvir}{\mbox{\object{QS\,Vir}}}

\newcommand{\ten}[2]{#1\times 10^{#2}}
\newcommand{\oc}{$O\!-\!C$}
\newcommand{\oclin}{$O\!-\!C_\mathrm{lin}$}
\newcommand{\oci}{$O\!-\!C_\mathrm{i}$}
\newcommand{\oco}{$O\!-\!C_\mathrm{o}$}
\newcommand{\ocell}{$O\!-\!C_\mathrm{ell}$}
\newcommand{\ei}{$e_\mathrm{i}$}
\newcommand{\eo}{$e_\mathrm{o}$}

\newcommand{\pou}{$P_\mathrm{o}$}
\newcommand{\pin}{$P_\mathrm{i}$}
\newcommand{\pbin}{$P_\mathrm{bin}$}
\newcommand{\mjup}{$M_\mathrm{Jup}$}
\newcommand{\chisq}{$\,\chi^2$}

\begin{document}

\title{The quest for companions to post-common envelope binaries}
\subtitle{IV. The 2:1 mean-motion resonance of the planets orbiting NN~Serpentis}

\author{
Beuermann, K. % \inst{1} 
\and 
Dreizler, S. % \inst{1} 
\and 
Hessman, F.~V. % \inst{1} 
} 

\institute{
Institut f\"ur Astrophysik, Georg-August-Universit\"at, Friedrich-Hund-Platz 1, D-37077 G\"ottingen, Germany 
}
\date{Received 7 October 2012; accepted 13 May 2013}

\authorrunning{K. Beuermann et al.} 
%\titlerunning{The quest for companions to post-common envelope binaries IV}
\titlerunning{The 2:1 mean-motion resonance of the planets orbiting NN~Serpentis}

\abstract{We present 69 new mid-eclipse times of the young post-common
  envelope binary (PCEB) \nnser, which was previously suggested to
  possess two circumbinary planets.  We have interpreted the observed
  eclipse-time variations in terms of the light-travel time effect caused by
  two planets, exhaustively covering the multi-dimensional parameter
  space by fits in the two binary and ten orbital parameters. We
  supplemented the fits by stability calculations for all models with
  an acceptable \chisq. An island of secularly stable 2\,:\,1 resonant
  solutions exists, which coincides with the global \chisq\
  minimum. Our best-fit stable solution yields current orbital periods
  $P_\mathrm{o}\!=\!15.47$\,yr and
  $P_\mathrm{i}\!=\!\,7.65$\,yr and eccentricities
  $e_\mathrm{o}\!=\!0.14$ and $e_\mathrm{i}\!=\!0.22$ for
  the outer and inner planets, respectively.  The companions qualify as
  giant planets, with masses of 7.0\,\mjup\ and 1.7\,\mjup\ for the
  case of orbits coplanar with that of the binary. The two-planet
  model that starts from the present system parameters has a lifetime
  greater than $10^8$\,yr, which significantly exceeds the age of NN~Ser
  of $10^6$\,yr as a PCEB. The resonance is characterized by libration
  of the resonant variable $\Theta_1$ and circulation of
  $\omega_\mathrm{i}\!-\!\omega_\mathrm{o}$, the difference between
  the arguments of periapse of the two planets.  No stable nonresonant
  solutions were found, and the possibility of a 5\,:\,2 resonance
  suggested previously by us is now excluded at the 99.3\% confidence
  level. }


\maketitle

%________________________________________________________________

\section{Introduction}

Planets orbiting post-common envelope binaries (PCEB) are a recent
discovery, and only a few PCEB are known or suspected to harbor
planetary systems. These planets are detected by the light-travel time
(LTT) effect, which measures the variations in the observed
mid-eclipse times caused by the motion of the binary about the common
center of mass. The derived orbital periods significantly exceed those
of planets orbiting single stars, because the LTT method preferably
selects long orbital periods and the radial-velocity and transit
methods short ones. Since eclipse-time variations can also
be brought about by other mechanisms, it is necessary to prove the
strict periodicity of the LTT signal to confirm its planetary
origin. The orbital periods on the order of a decade represent a
substantial challenge, however. For a system of circumbinary planets,
one additionally has to demonstrate the secular stability of a
particular solution.

The discovery of two planets orbiting the dG/dM binary Kepler~47 with
semi-major axes less than 1\,AU \citep{oroszetal12} has convincingly
demonstrated that close binaries can possess planetary systems, but
also raised questions about their orbital co-evolution.  The
evolutionary history of planets orbiting PCEB may be complex: they
either formed before the common-envelope (CE) event and had to survive
the loss of a substantial amount of matter from the evolving binary
and the passage through the ejected shell, or they were assembled
later from CE-material. Even if the planets existed before the CE, their
masses may have increased in the CE by accretion, making a distinction
between first and second-generation origins difficult. In both cases
the CE may have significantly affected the planetary orbits, and we
expect that the dynamical age of the system equals the age of the
PCEB, which is the cooling age of the white dwarf. The hot white dwarf
in \nnser\ has an age of only $10^6$\,yr \citep{parsonsetal10a}, so
the planets in \nnser\ are dynamically young. Migration of planets is
expected to occur in the CE, but our poor knowledge of the CE
structure complicates predictions about the dynamical state of PCEB
planetary systems.

Only a handful of PCEB have been suggested to contain more than one
circumbinary companion. Of these, \nnser\
\citep{beuermannetal10,horneretal12} and \hwvir\ \citep{beuermannetal12}
have passed the test of secular stability. A final conclusion on
\huaqr\ \citep{gozdziewskietal12,hinseetal12} is pending, and the
eclipse-time variations in \qsvir\ are presently not understood
\citep{parsonsetal10b}. For a few other contenders, the data are still
insufficient for theoretical modeling. In the case of \nnser, a period
ratio of the two proposed planets near either 2\,:\,1 or 5\,:\,2 was
found (Beuermann et al. 2010, henceforth Paper~I), with both models
being stable for more than $10^6$\,yr.  \citet{horneretal12} confirm
the dynamical stability of the proposed orbits, but suggest that
further observations are vital in order to better
constrain the system's true architecture. In this paper, we report
further eclipse-time observations of \nnser\ and present the results
of extensive stability calculations, which show that a two-planet
system that starts from our best fit will be secularly stable and stay
tightly locked in the 2\,:\,1 mean-motion resonance for more than $10^8$\,yr.

%\vspace*{-3mm}
\section{Observations}
\label{sec:obs}

\begin{table}[t]
\begin{flushleft}
\caption{Mid-eclipse times of \nnser\ measured with MONET/N.}
\begin{tabular}{lcccc}
\hline \hline\\[-1ex]
Cycle   &  JD         & Error & BJD(TDB)     & \ocell\ \\
        & 2450000+   &  (days) & 2450000+    & (s)\\[0.5ex]
\hline\\[-1ex]
\multicolumn{5}{l}{\emph{(a) January /February 2010 (Paper~I). }}\\[0.5ex]
60489 & 5212.9431997 & 0.0000069 &  5212.9418190 & 0.20 \\
60505 & 5215.0243170 & 0.0000067 &  5215.0230965 & \hspace{-2mm}$-$0.17 \\
60528 & 5218.0159241 & 0.0000044 &  5218.0149383 & \hspace{-2mm}$-$0.23 \\
60735 & 5244.9402624 & 0.0000029 &  5244.9415257 &  0.09 \\[0.7ex]
60743 & 5245.9808144 & 0.0000032 &  5245.9821657 &  0.02 \\ 
60751 & 5247.0213676 & 0.0000034 &  5247.0228067 &  0.02 \\
60774 & 5250.0129571 & 0.0000034 &  5250.0146473 & \hspace{-2mm}$-$0.16 \\[1.4ex]
\multicolumn{5}{l}{\emph{(b) September 2010  to February 2013 (this work). }}\\[0.7ex]
62316 & 5450.5995341 & 0.0000052 &  5450.5981913 &  \hspace{-2mm}$-$0.35\\  
62339 & 5453.5916006 & 0.0000067 &  5453.5900372 &  \hspace{-2mm}$-$0.10\\  
62347 & 5454.6323120 & 0.0000067 &  5454.6306732 &  \hspace{-2mm}$-$0.53\\  
62462 & 5469.5925302 & 0.0000049 &  5469.5899042 &   0.87\\  
62531 & 5478.5685430 & 0.0000051 &  5478.5654275 &   0.39\\[0.7ex]  
63403 & 5591.9955682 & 0.0000030 &  5591.9953015 &   0.20\\  
63449 & 5597.9787495 & 0.0000031 &  5597.9789848 &  \hspace{-2mm}$-$0.06\\  
63457 & 5599.0193095 & 0.0000027 &  5599.0196329 &   0.55\\  
63472 & 5600.9703359 & 0.0000067 &  5600.9708248 &  \hspace{-2mm}$-$0.33\\[0.7ex]  
63671 & 5626.8541370 & 0.0000041 &  5626.8567788 &   0.21\\  
63672 & 5626.9842067 & 0.0000037 &  5626.9868588 &   0.20\\  
63679 & 5627.8946901 & 0.0000041 &  5627.8974134 &  \hspace{-2mm}$-$0.35\\  
63740 & 5635.8289827 & 0.0000044 &  5635.8323043 &  \hspace{-2mm}$-$0.15\\  
63741 & 5635.9590539 & 0.0000049 &  5635.9623850 &  \hspace{-2mm}$-$0.10\\  
63756 & 5637.9101120 & 0.0000042 &  5637.9135830 &  \hspace{-2mm}$-$0.45\\[0.7ex]  
63833 & 5647.9256179 & 0.0000033 &  5647.9297557 &  \hspace{-2mm}$-$0.30\\  
63864 & 5651.9578679 & 0.0000046 &  5651.9622470 &   0.30\\  
63879 & 5653.9089538 & 0.0000031 &  5653.9134435 &  \hspace{-2mm}$-$0.19\\  
63886 & 5654.8194618 & 0.0000037 &  5654.8240016 &  \hspace{-2mm}$-$0.44\\  
63925 & 5659.8923285 & 0.0000038 &  5659.8971306 &  \hspace{-2mm}$-$0.14\\  
63933 & 5660.9329164 & 0.0000053 &  5660.9377686 &  \hspace{-2mm}$-$0.42\\[0.7ex]  
64079 & 5679.9239504 & 0.0000038 &  5679.9294753 &   0.08\\  
64086 & 5680.8344900 & 0.0000039 &  5680.8400351 &  \hspace{-2mm}$-$0.02\\  
64116 & 5684.7368230 & 0.0000052 &  5684.7424416 &   0.17\\  
64132 & 5686.8180703 & 0.0000035 &  5686.8237195 &  \hspace{-2mm}$-$0.22\\[0.7ex]   
64784 & 5771.6337400 & 0.0000040 &  5771.6359752 &  \hspace{-2mm}$-$0.30\\  
64869 & 5782.6914476 & 0.0000067 &  5782.6927870 &  \hspace{-2mm}$-$0.40\\  
64938 & 5791.6677191 & 0.0000043 &  5791.6683161 &  \hspace{-2mm}$-$0.52\\  
64961 & 5794.6598199 & 0.0000041 &  5794.6601694 &   0.33\\  
64976 & 5796.6111770 & 0.0000036 &  5796.6113657 &  \hspace{-2mm}$-$0.19\\[1.3ex] 
\hline\\[-5ex]                                                              
\end{tabular}
\label{tab:monet}
\end{flushleft}
\end{table}

\begin{table}[t]
\begin{flushleft}

\vspace{1.5mm}
{\bf Table 1 continued}

\vspace{4.0mm}
\begin{tabular}{lcccc}
\hline\hline \\[-1ex]
Cycle   &  JD         & Error & BJD(TDB)     & \ocell\ \\
        & 2450000+   & (days) & 2450000+    & (s)\\[0.5ex]
\hline\\[-1ex]
64992 & 5798.6926278 & 0.0000045 &  5798.6926456 & \hspace{-2mm}$-$0.42\\[0.64ex]  
65053 & 5806.6281610 & 0.0000047 &  5806.6275377 & \hspace{-2mm}$-$0.18\\  
65081 & 5810.2706975 & 0.0000040 &  5810.2697878 &  0.32\\  
65084 & 5810.6609723 & 0.0000097 &  5810.6600323 &  0.67\\  
65099 & 5812.6123145 & 0.0000027 &  5812.6112242 & \hspace{-2mm}$-$0.23\\  
65099 & 5812.6123171 & 0.0000048 &  5812.6112268 & \hspace{-2mm}$-$0.01\\[0.64ex]  
65360 & 5846.5653892 & 0.0000036 &  5846.5621492 &  0.13\\  
65460 & 5859.5738976 & 0.0000089 &  5859.5701607 & \hspace{-2mm}$-$0.24\\[0.64ex]  
65963 & 5925.0031268 & 0.0000080 &  5925.0004922 &  0.66\\  
65994 & 5929.0353491 & 0.0000050 &  5929.0329653 & \hspace{-2mm}$-$0.38\\  
66209 & 5957.0004918 & 0.0000035 &  5957.0002090 &  0.32\\[0.64ex]  
66324 & 5971.9584470 & 0.0000038 &  5971.9594273 &  0.25\\  
66332 & 5972.9990053 & 0.0000056 &  5973.0000740 &  0.71\\  
66362 & 5976.9010792 & 0.0000080 &  5976.9024783 &  0.65\\  
66370 & 5977.9416245 & 0.0000044 &  5977.9431113 & \hspace{-2mm}$-$0.07\\  
66409 & 5983.0143233 & 0.0000047 &  5983.0162339 & \hspace{-2mm}$-$0.42\\  
66416 & 5983.9248143 & 0.0000036 &  5983.9268000 &  0.01\\[0.64ex]  
66615 & 6009.8088279 & 0.0000045 &  6009.8127551 &  0.12\\  
66631 & 6011.8899714 & 0.0000029 &  6011.8940322 & \hspace{-2mm}$-$0.36\\  
66669 & 6016.8327239 & 0.0000032 &  6016.8370844 &  0.14\\  
66670 & 6016.9627977 & 0.0000028 &  6016.9671658 &  0.24\\  
66677 & 6017.8733096 & 0.0000033 &  6017.8777301 &  0.50\\  
66685 & 6018.9138903 & 0.0000027 &  6018.9183694 &  0.33\\[0.64ex]  
66815 & 6035.8235355 & 0.0000061 &  6035.8287909 &  0.25\\  
66893 & 6045.9694909 & 0.0000055 &  6045.9750363 & \hspace{-2mm}$-$0.44\\  
66900 & 6046.8800348 & 0.0000041 &  6046.8855994 & \hspace{-2mm}$-$0.28\\  
66908 & 6047.9206573 & 0.0000031 &  6047.9262424 & \hspace{-2mm}$-$0.14\\[0.64ex]  
67284 & 6096.8315332 & 0.0000034 &  6096.8363914 &  0.13\\  
67330 & 6102.8155143 & 0.0000035 &  6102.8200727 & \hspace{-2mm}$-$0.46\\  
67337 & 6103.7261262 & 0.0000028 &  6103.7306356 & \hspace{-2mm}$-$0.32\\  
67352 & 6105.6774387 & 0.0000036 &  6105.6818401 & \hspace{-2mm}$-$0.16\\[0.64ex] 
67675 & 6147.6963688 & 0.0000058 &  6147.6977420 &  0.19\\  
67698 & 6150.6884525 & 0.0000041 &  6150.6895788 & \hspace{-2mm}$-$0.45\\  
67775 & 6160.7054661 & 0.0000046 &  6160.7057628 &  0.42\\[0.64ex]  
67928 & 6180.6093127 & 0.0000034 &  6180.6080271 &  0.11\\  
67936 & 6181.6500361 & 0.0000036 &  6181.6486726 &  0.46\\  
68028 & 6193.6182396 & 0.0000047 &  6193.6160366 & \hspace{-2mm}$-$0.65\\[0.64ex]   
69168 & 6341.9060724 & 0.0000028 &  6341.9074599 &  0.06\\ [0.64ex]        
69575 & 6394.8450806 & 0.0000020 &  6394.8500987 &  0.04\\ [1.2ex]                
\hline\\[-5ex]                                                              
\end{tabular}
\end{flushleft}
\end{table}

We obtained 69 additional mid-eclipse times of the 16.8\,mag binary
with the MONET/North telescope at the University of Texas' McDonald
Observatory via the MONET browser-based remote-observing
interface. The photometric data were taken with an Apogee ALTA E47+
1k$\times$1k CCD camera mostly in white light with 10\,s exposures
separated by 3\,s readout. In most cases photometry was performed
relative to a comparison star 2.0\,arcmin SSW of \nnser, but in some
nights absolute photometry yielded lower uncertainties. The eclipse
light curves were analyzed with the white dwarf represented by a
uniform disk occulted by the secondary star
\citep[see][]{backhausetal12}. A seven-parameter fit was made to each
individual eclipse profile of the relative or absolute source
flux. The free parameters of the fit were the mid-eclipse time
$T_\mathrm{c}$, the fluxes outside and inside eclipse, the FWHM of the
profile, the ingress/egress time, and the two parameters $a_1$ and
$a_2$ of a multiplicative function
$f\!=\!1+a_1(t-T_\mathrm{c})+a_2(t-T_\mathrm{c})^2$, which allowed us
to model photometric variations outside of the eclipse. Such
variations can be caused by effects intrinsic to the source, as the
illumination of the secondary star, or by observational effects, as
color-dependent atmospheric extinction. The formal \mbox{1-$\sigma$}
error of $T_\mathrm{c}$ was calculated from the measurement errors of
the fluxes in the individual CCD images, employing standard error
propagation.  The distribution of the measured FWHM of all eclipses is
Gaussian with a mean of 569.20\,s and a standard deviation of
0.79\,s. In columns 1--3 of Table~\ref{tab:monet} we list the cycle
numbers, the observed mid-eclipse times $T_\mathrm{c}$ in UTC, and the
1-$\sigma$ errors of the 69 new eclipses.  Re-analysis of earlier
MONET/N data led to small corrections, and we also include the seven
mid-eclipse times published already in Paper~I. The errors of
$T_\mathrm{c}$ vary between 0.17 and 0.83\,s, depending on the quality
of the individual light curves. The mean timing error is 0.38\,s. All
mid-eclipse times were converted from Julian days (UTC) to barycentric
dynamical time (TDB) and corrected for the light travel time to the
solar system barycenter, using the tool provided by
\citet{eastmanetal10}\footnote{http://astroutils.astronomy.ohio-state.edu/time/}. 
These corrected times are given as barycentric Julian days in TDB in
column~4 of Table~\ref{tab:monet}. Together with the errors in column
3, they represent the new data subjected to the LTT fit in the
Section~\ref{sec:fit}.  

% Fig. 1
\begin{figure}[t]
\includegraphics[bb=120 10 510 680,height=89mm,angle=-90,clip]{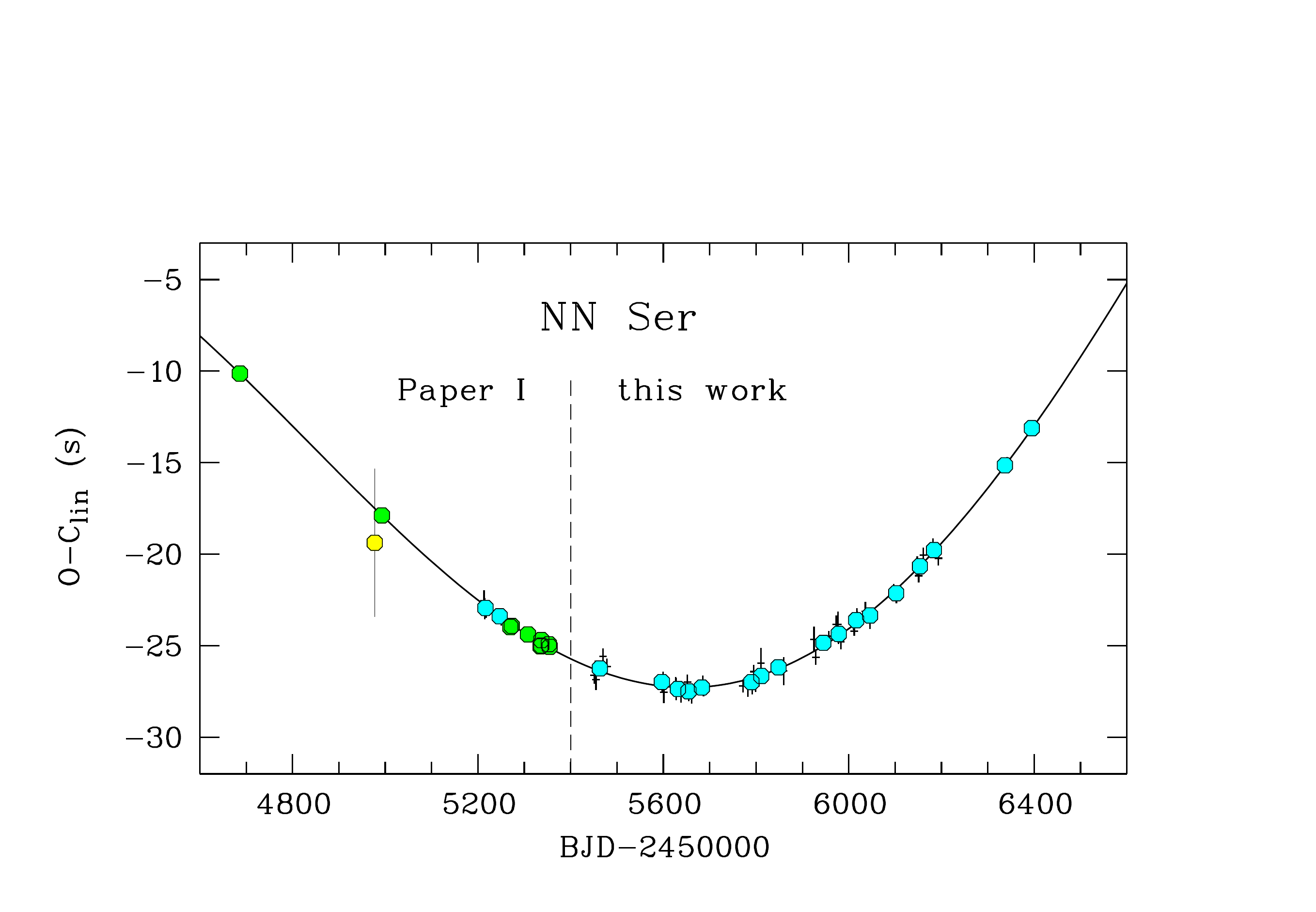}
\caption{Residuals of the mid-eclipse times of \nnser\ since 2007
  relative to the linear ephemeris of Eq.~1, with the best-fitting
  model of Fig.~\ref{fig:oc1} and Table~\ref{tab:nnser} shown as solid
  curve. See text for further explanation.}
\label{fig:new}
\end{figure}

Figure~\ref{fig:new} shows the \oclin\ residuals relative to the
linear ephemeris of the binary quoted in Eq.~\ref{eq:ephem} and
derived in Sect.~\ref{sec:fit} below. The data points to the left of
the dashed line are from Paper~I and those to the right from this
work. The residuals of the individual MONET eclipse times are
displayed as crosses with error bars. Overplotted are the weighted
mean \oc\ values for 19 groups of timings typically collected in the
dark periods of individual months (cyan-blue filled circles). Plotting
these mean values avoids cluttering up the graphs on the expanded
ordinate scales in Figs.~\ref{fig:oc1} and \ref{fig:oc2}.
All fits presented in this paper were made to the total set of 121
individual mid-eclipse times, 52 from Paper~I and 69 from this work.
The solid curve in Fig.~\ref{fig:new} represents the best-fit
two-planet LTT model derived in Sect.~\ref{sec:fit}. The residuals
\ocell\ of the individual timings relative to this fit are included in
column~5 of Table~\ref{tab:monet}. The new feature that allows us to
derive a significantly improved orbital solution within the framework
of the two-planet LTT model is the detection of the upturn in \oclin\
that commences near JD\,2455650.

\section{General approach}

In this paper, we adopt a purely planetary explanation of the
eclipse-time variations in \nnser\ and  represent the set of
mid-eclipse times by the sum of the linear ephemeris of
the binary and the LTT effect of two planets. A model that only 
involves a single planet was already excluded in Paper~I and is not
discussed again, given the very poor fit with a reduced
$\,\chi^2_\nu\!=\!40.3$. Specifically, we describe the motion of the
center of mass of the binary in barycentric coordinates by the
superposition of two Kepler ellipses, which reflect the motion of the
planets (Irwin, 1952, Eq.~1; Kopal, 1959, Eq. 8-94, Beuermann et al.,
2012, Eq.~2; Paper I, Eq.~1\footnote{ Eq.~1 of Paper~I contains a
  misprinted sign in the last bracket.}).  We justify the Keplerian
model at the end of this section.

The central binary was treated as a single object with the combined
mass of the binary components $M_\mathrm{bin}\!=\!0.646$\,$M_\odot$,
\citep{parsonsetal10a}. This approach is justified, because the
gravitational force exerted by the central binary on either of the
planets varies only by a fraction of $10^{-7}$ or less over the binary
period. Hence, the \nnser\ system was treated as a triple consisting
of the central object and two planets.  We assumed coplanar planetary
orbits viewed edge-on, which coincide practically with the orbital
plane of the binary with an inclination $i_\mathrm{bin}\!=\!89.6^\circ\pm 0.2^\circ$
\citep{parsonsetal10a}. In spite of the high inclination, transit
events, which would present proof of the planetary hypothesis, are
unfortunately extremely improbable. Accurate eclipse-time measurements
over a sufficiently large number of orbits could, in principle,
provide information on the inclinations of the gravitationally
interacting orbits, but this is presently not feasible given the long
periods of the planets in \nnser.  The measured orbital periods and
amplitudes of the LTT effect yield minimum masses, with the true
masses scaling as 1/sin\,$i_\mathrm{k}$, where $i_\mathrm{k}$ is the
unknown inclination of planet $k$. The semi-major axes depend on the
total system mass, implying a minor dependence on~$i_\mathrm{k}$.

% Fig. 2
\begin{figure}[t]
\includegraphics[bb=120 50 520 720,height=89mm,angle=-90,clip]{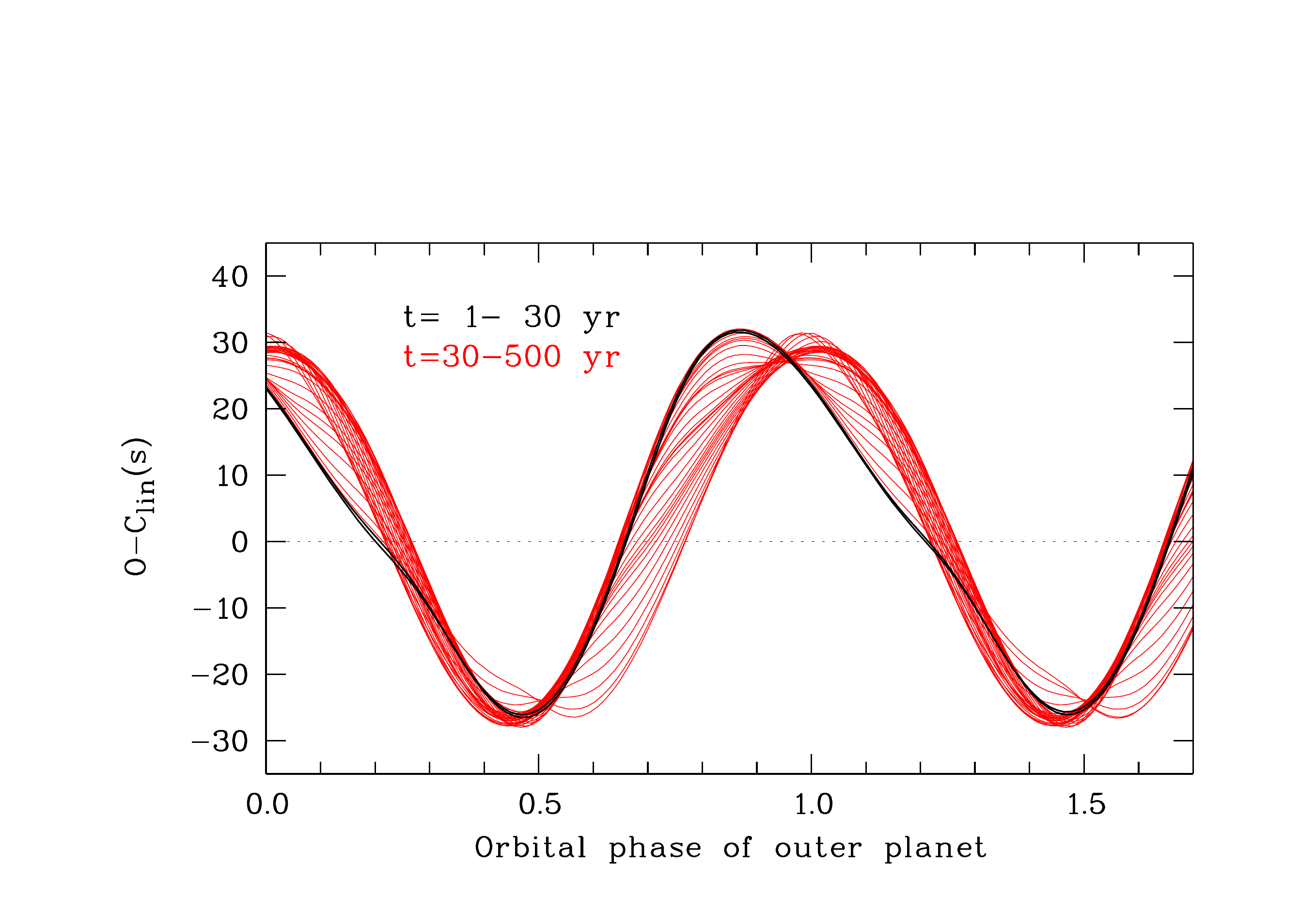}
\caption[chart]{Temporal variation of the mid-eclipse times of \nnser\
  as predicted by our best-fitting dynamical two-planet model of
  Table~\ref{tab:nnser} relative to the underlying linear ephemeris of
  the binary and folded over the orbital period of the outer
  planet. The first two orbits, covering a time interval of 30\,yr
  (black curves), and the next 29 orbits (30--500\,yr, red curves) are
  displayed. }
\label{fig:kepler}
\end{figure}

Even with the new data, the least-squares fits of the LTT model permit
a wide range of model parameters.
As in Paper~I, we required, therefore, that an acceptable model
provides a good fit to the data \emph{and} fulfills the side condition
of secular stability.  Formally, a minimum lifetime of only $10^6$\,yr
is required, the cooling age of the white dwarf and the age of \nnser\
as a PCEB, but most models in the vicinity of the best fit reached
more than $10^8$\,yr, suggesting that a truly secularly stable
solution exists.

% Fig. 3
\begin{figure*}[t]
% \hspace{11.90mm} \includegraphics[bb=35 94 546 772,width=50.5mm,clip]{20510f3a.eps}
% \hspace{4.90mm} \includegraphics[bb=35 94 546 772,width=50.5mm,clip]{20510f3b.eps}
% \hspace{4.90mm} \includegraphics[bb=35 94 546 772,width=50.5mm,clip]{20510f3c.eps}
% \vspace{-71.1mm} 
%
% \includegraphics[bb=48  36 518 487,width=77.0mm,angle=-90,clip]{20510f3d.ps}
% \hspace{2mm}
% \includegraphics[bb=48 108 518 487,width=77.0mm,angle=-90,clip]{20510f3e.ps}
% \hspace{2mm}
% \includegraphics[bb=48 108 518 520,width=77.0mm,angle=-90,clip]{20510f3f.ps}
\includegraphics[width=18cm]{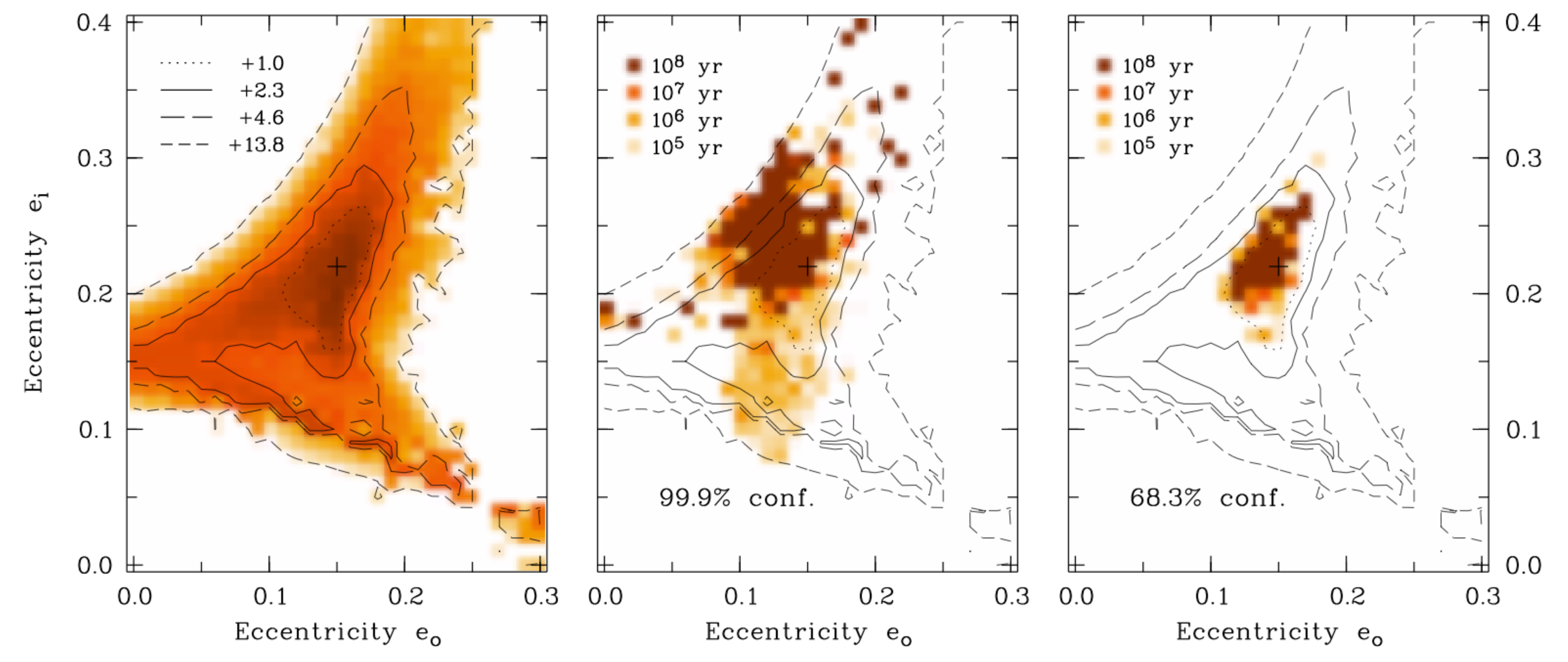}
\caption[chart]{Results of fitting the Keplerian two-planet model to the
  eclipse-time data of \nnser. \emph{Left: } \,$\chi^2$ distribution
  in the \eo,\ei\ plane. The + sign indicates the minimum \,$\chi^2$ and
  the contour lines refer to the increments  $\Delta \,\chi^2$ indicated in the
  legend. \emph{Center: } Maximum lifetime $\tau$ of models with
  $\chi^2$ at the 99.9\% confidence level ($\Delta
  \,\chi^2\!=\!+13.8$ for two degrees of freedom).  For ease of comparison, the \,$\chi^2$
  contours of the left panel are included. \emph{Right: } Same for
  models with  $\chi^2$ at the 68.3\% confidence level ($\Delta \,\chi^2\!=\!+2.3$).
}
\label{fig:gray}
\end{figure*}

% Fig. 4
\begin{figure*}[t]
\includegraphics[bb=201  33 524 449,width=49.0mm,angle=-90,clip]{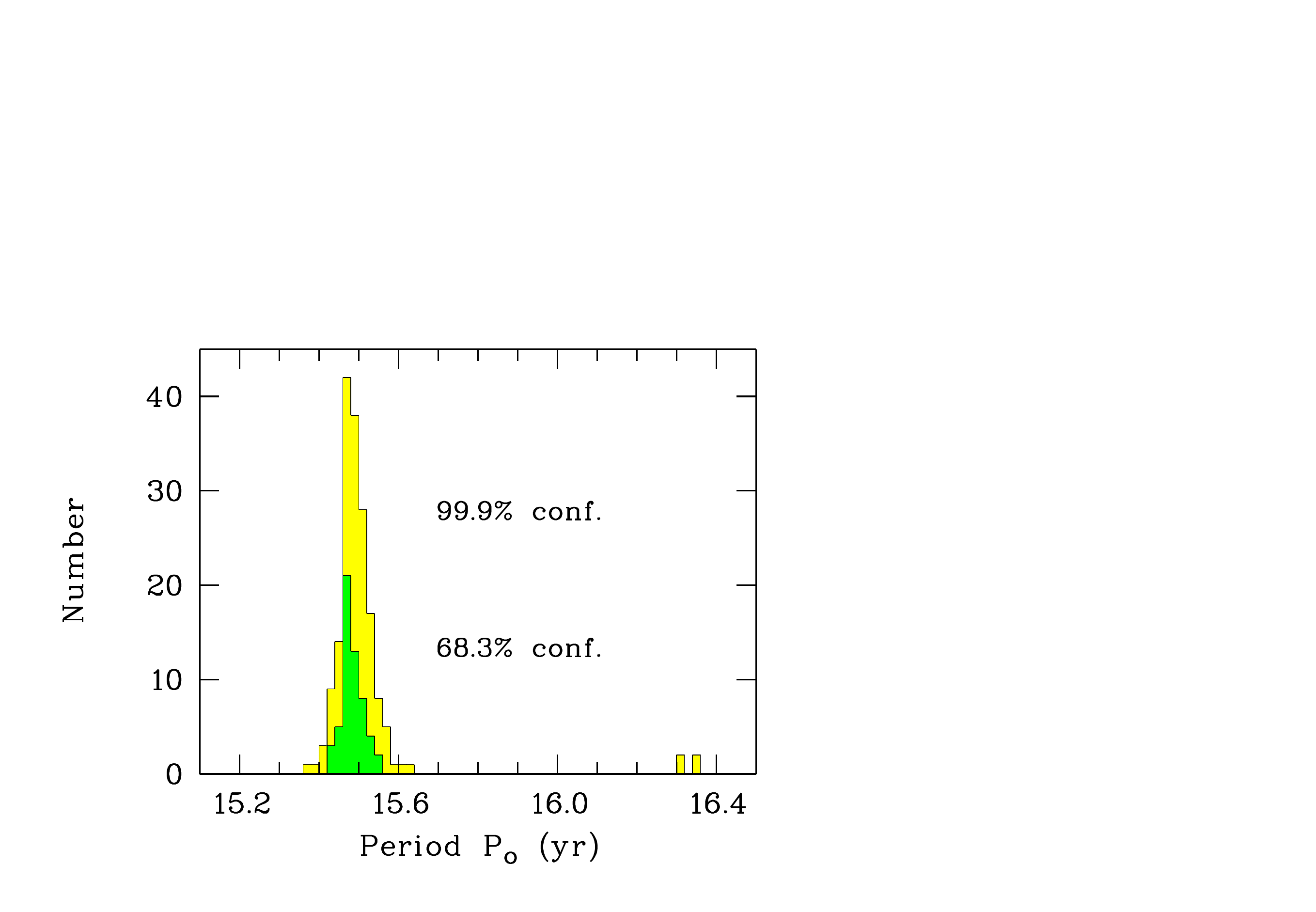} \hspace{2.90mm}
\includegraphics[bb=201 113 524 449,width=49.0mm,angle=-90,clip]{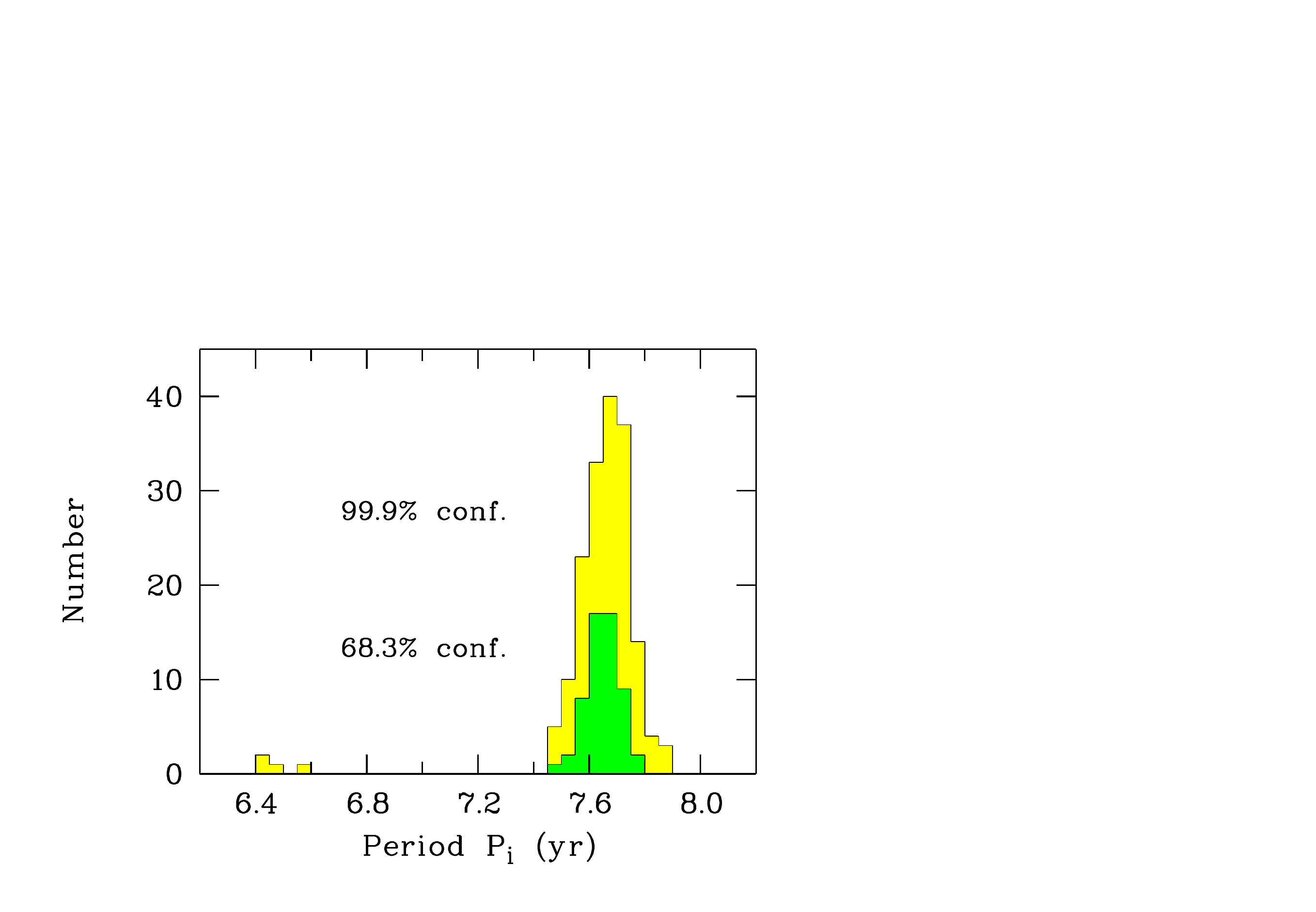} \hspace{2.90mm}
\includegraphics[bb=201 113 524 482,width=49.0mm,angle=-90,clip]{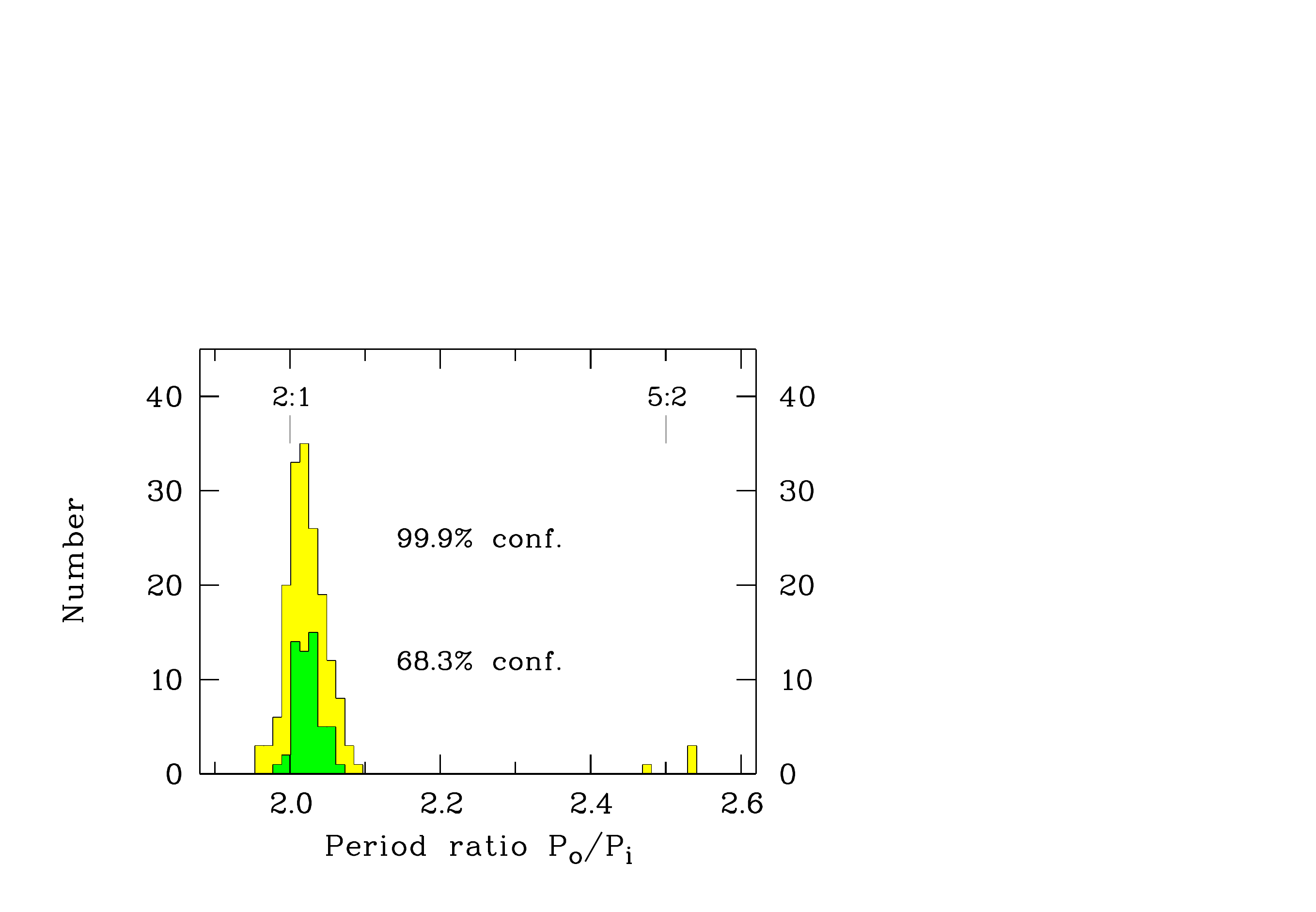}
\caption[chart]{Histograms of the orbital periods \pou\ and \pin\ of
  the outer and inner planets of \nnser\ and the period ratio
  $P_\mathrm{o}/P_\mathrm{i}$for solutions with a lifetime exceeding
  $10^6$\,yr and a $\,\chi^2$ corresponding to the 99.9\% confidence
  level (yellow) or the 68.3\% confidence level (green) for two
  degrees of freedom.}
\label{fig:prat}
\end{figure*}

We performed a large number of least-squares fits in search of the
global $\,\chi^2$ minimum, using the Levenberg-Marquardt minimization
algorithm implemented in {\tt mpfit} of IDL
\citep{marquardt63,markwardt09}. The time evolution of all models that
achieved a $\,\chi^2$ below a preset limit (Sect.~\ref{sec:fit}) was
followed numerically until they became either unstable or reached a
lifetime of $10^8$\,yr.  Most models that fit the data developed an
instability within a few hundred years or less and fewer than 200
survived for $10^8$\,yr, allowing us to calculate the evolution of all
models with an acceptable overall CPU-time requirement. We used the
hybrid symplectic integrator in the {\tt mercury6} package
\citep{chambers99}, which allows one to evolve planetary
systems very efficiently with high precision over long times. The
model treats the central object and the planets as point masses and
angular momentum and energy are conserved to a high degree of
accuracy. Time steps of 0.1\,yr were used, which is not adequate for
the treatment of close encounters, but such incidents do not occur in
the successful models (see Beuermann et al. 2012 for more details).

These dynamical model calculations provide us with information on the
complex time variations of the orbital parameters, but also allow us
to test the validity of the Keplerian assumption in fitting the
data. Figure~\ref{fig:kepler} shows the calculated variation of the
mid-eclipse times for the first 500\,yr of our best-fit longlived
model, starting from the Keplerian fit. The LTT effect is displayed
folded over the 15.47-yr orbital period of the outer planet, which
contributes 87\% of the signal. The first two orbits (30 yr) agree
closely and are indistinguishable from the Keplerian fit, but in the
long run the signal changes due to the dynamical evolution. The
orbital phase interval covered in Fig.~\ref{fig:kepler} is the same as
in Fig~\ref{fig:oc1} (top panel), where the data and the Keplerian fit
are displayed. We conclude that in this special case fitting the data
by a Keplerian model is justified. Data trains that extend over more
orbits or involve more massive companions will
require a dynamical model.

\section{Results}
\label{sec:results}

Our least-squares fit to the 121 mid-eclipse times involves twelve
free parameters, two for the linear ephemeris of the binary, the epoch
$T_\mathrm{bin}$ and period \pbin, and five orbital elements for each
planet. With all parameters free, the number of degrees of
freedom of the fit is therefore 109. The five parameters for planet $k$
are the orbital period $P_\mathrm{k}$, the eccentricity
$e_\mathrm{k}$, the argument of periapse $\omega_\mathrm{bin,k}$
measured from the ascending node in the plane of the sky, the time
$T_\mathrm{k}$ of periapse passage, and the amplitude of the eclipse
time variation,
$K_\mathrm{bin,k}=a_\mathrm{bin,k}$\,sin\,$i_\mathrm{k}$/c, with
$a_\mathrm{bin,k}$ the semi-major axis of the orbit of the center of
mass of the binary about the common center of mass of the system,
$i_\mathrm{k}$ the inclination of the planet's orbit against the plane
of the sky, and c the speed of light. The arguments of periapse of the
center of mass of the binary (with the index `bin') and that of the
planet differ by $\pi$.  For clarity, we use the indices k=i or k=o
for the inner and outer planets, respectively\,\footnote{There is no
  official nomenclature for the naming of exo\-planets. At the time of
  Paper~I, the A\&A editor considered assigning the first planet
  discovered around a binary the small letter `b' as illogical and
  suggested that the two planets orbiting the binary \nnser\ should be
  named `c' and `d', `ab' being the binary components. In the
  nomenclature proposed by \citet{hessmanetal10}, the planets in
  \nnser\ would be labeled (AB)b and (AB)c. Since neither convention
  has been officially adopted, we prefer the neutral designations
  `outer' and `inner planet', which avoids any misunderstanding, as
  long as there are only two planets.}.

We fitted a total of 101,618 models to the data, adopting three lines
of approach. Run~1 with 81,552 models is a grid search in the \eo,\ei\
plane. In Run~2 with 10,598 models, we started again
from the \eo,\ei\ grid values, but subsequently optimized the
eccentricities along with the other parameters. Run~3 with 9,468
models was performed to study the properties of selected solutions, in
particular, models in the vicinity of the best fit. It also includes
models with a circular outer orbit, as advocated in Paper~I, but
dismissed now and not discussed further in this paper.

\subsection{Grid search in the $e_\mathrm{o},e_\mathrm{i}-$plane}
\label{sec:fit}

The grid search in the \eo,\ei\ plane covers eccentricities from zero
to 0.40 in steps of 0.01, with an extension to $e_\mathrm{o}\!=\!0.70$
in steps of 0.05. The ten other parameters were optimized, starting
from values chosen randomly within conservative limits.
Figure~\ref{fig:gray} (left panel) shows the resulting \chisq\
distribution for $e_\mathrm{o}\!<\!0.3$. For each grid point, the
figure displays the best \chisq\ value of typically 50 model fits. The
minimum is attained at $e_\mathrm{o},e_\mathrm{i}\!=\!0.15,0.22$ with
$\chi^2_\mathrm{min}\!=\!95.4$. In the extension of the grid to
$e_\mathrm{o}\!=\!0.70$, no model fit with $\chi^2\!<\!150$ was
found. The contour lines refer to increments $\Delta\chi^2$ of +1.0,
+2.3, +4.6, and +13.8, corresponding to confidence levels of 40\%,
68.3\%, 90.0\%, and 99.9\% for two degrees of freedom. Color coding
displays the lowest \chisq\ as dark red, fading into white at the
99.9\% level. The 68.3\% contour encloses a substantial fraction of
the \eo,\ei\ plane, indicating that the data permit fits with a wide
range of eccentricities and appropriate adjustment of the remaining
parameters. Notably, those describing the inner planet are not well
defined by the data alone, and the independent stability information is
required for further selection.

% Fig. 5
\begin{figure}[t]
\includegraphics[bb=179 25 493 679,height=89mm,angle=-90,clip]{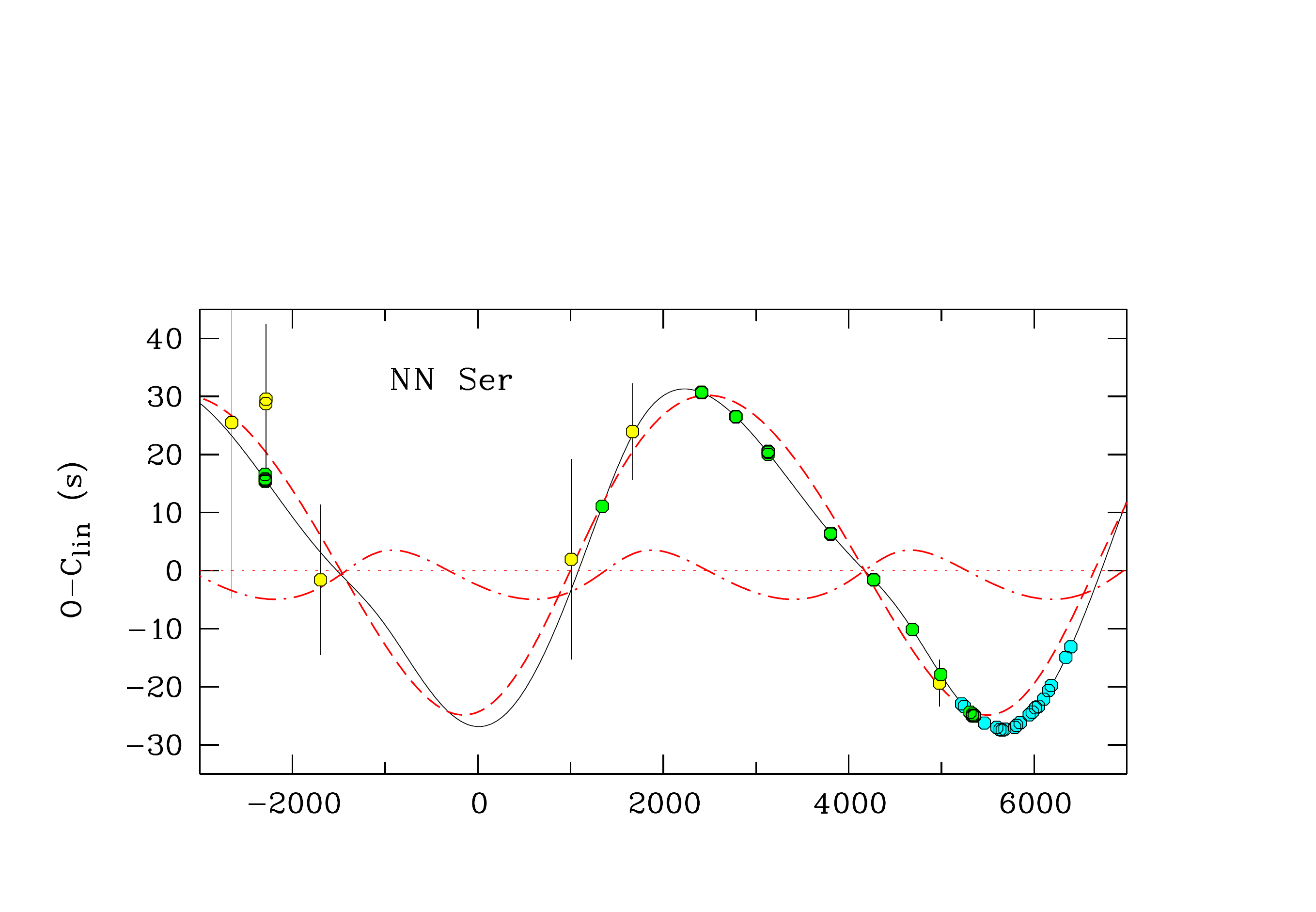}
\includegraphics[bb=335 25 493 679,height=89mm,angle=-90,clip]{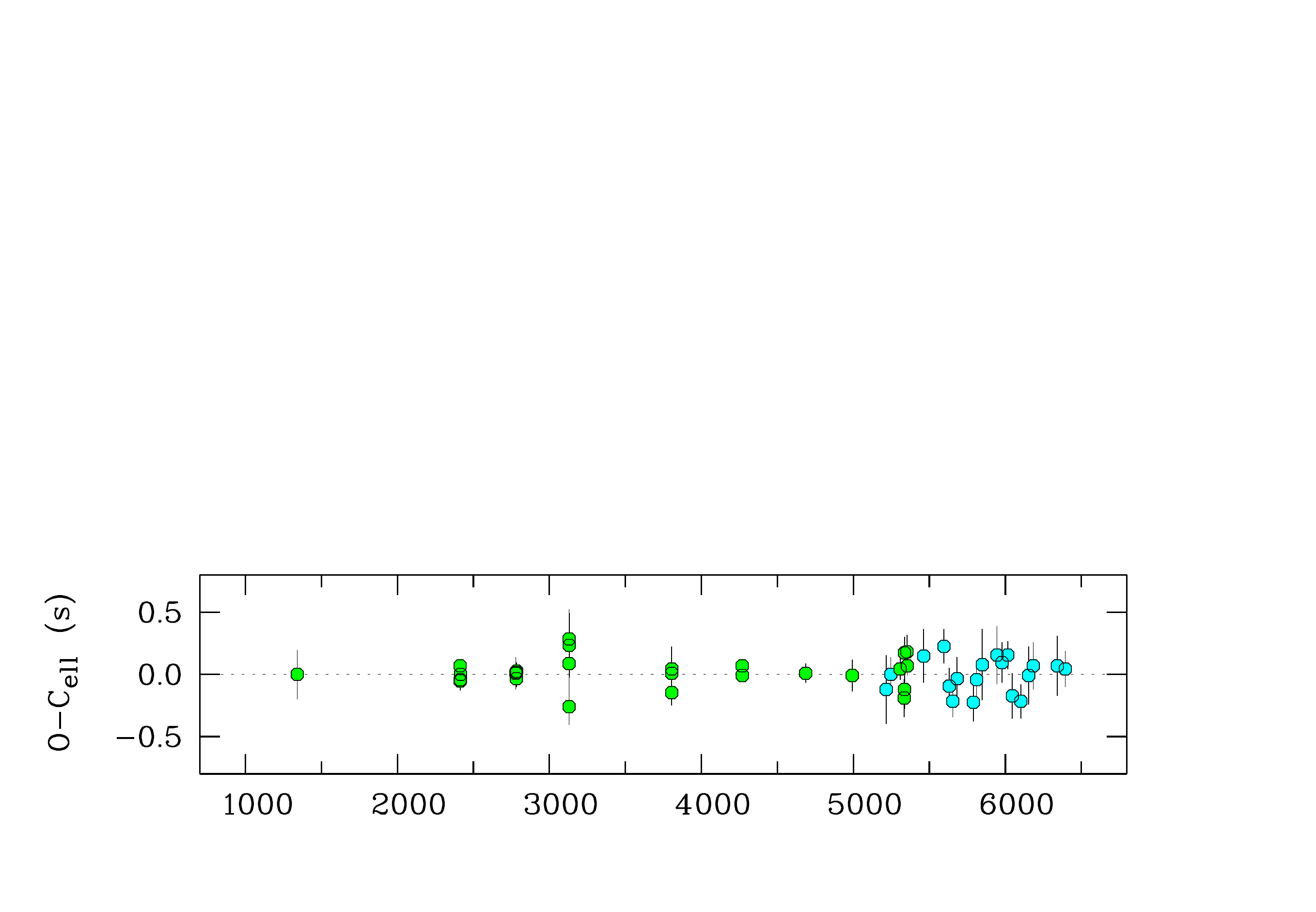}
\includegraphics[bb=335 25 510 679,height=89mm,angle=-90,clip]{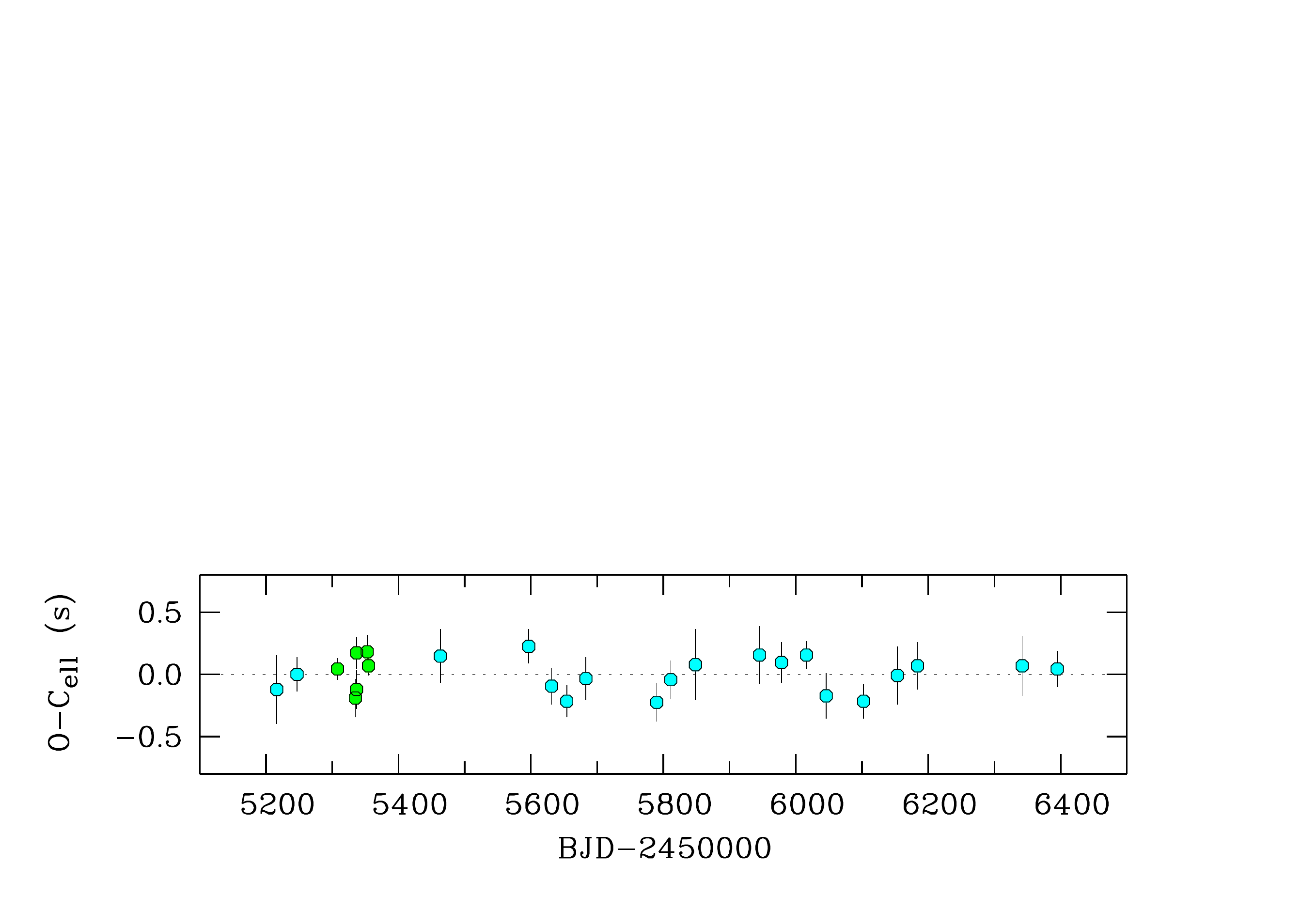}
\caption[chart]{ Best fit of the Keplerian two-planet model to the
  eclipse-time variations of \nnser.  The lower panels display the
  residuals from the fit for two different time intervals. Data points
  with errors larger than 1\,s are omitted. }
\label{fig:oc1}
\end{figure}

We calculated the temporal evolution of all models with
$\,\chi^2\!<\!110$ (99.93\% confidence level), starting from the fit
parameters and requiring that the lifetime $\tau$ exceeds the age of
\nnser\ of $10^6$\,yr. We found that an island of secularly stable
models is located close to the minimum $\,\chi^2$, establishing an
internal consistency between the fits to the data and the results of
the stability calculations. The requirement of $\tau\!>\!10^6$\,yr
imposes severe constraints on the orbital parameters of the successful
models and, not surprisingly, the minimum \chisq\ of the stable models
is slightly inferior to that of the unconstrained fits. The best
stable model has $\chi^2_\mathrm{min}\!=\!95.6$, eccentricities
$e_\mathrm{o},e_\mathrm{i}\!=\!0.14,0.21$, and a lifetime
$\tau\!=\!\ten{5}{6}$\,yr. The center panel of Fig.~\ref{fig:gray}
shows the lifetime distribution in the \eo,\ei-plane, with the peak
lifetime in each bin displayed. The detailed structure of the lifetime
distribution is complex. In particular, the appearance of
secularly stable solutions outside the main island of stability is
reminiscent of a skerry landscape. These solutions fit the data less
well and disappear if the \chisq\ limit is reduced or, to stay in the
picture, the sea level is raised. This is shown in the
righthand panel of Fig.~\ref{fig:gray}, where only solutions with
$\tau\!>\!10^6$\,yr and $\,\chi^2\!<\!97.9$ are retained (68.3\%
confidence). The efficiency of the lifetime selection is demonstrated
by the reduced size of the island of stability, which covers only
a fraction of the $\,\chi^2$-space delineated by the 1-$\sigma$
contour level in the lefthand panel (solid curve). Its size decreases a
bit further if the eccentricities are also optimized as done in
Run~2.

% Fig. 6
\begin{figure}[t]
\includegraphics[bb=168 29 461 689,height=89mm,angle=-90,clip]{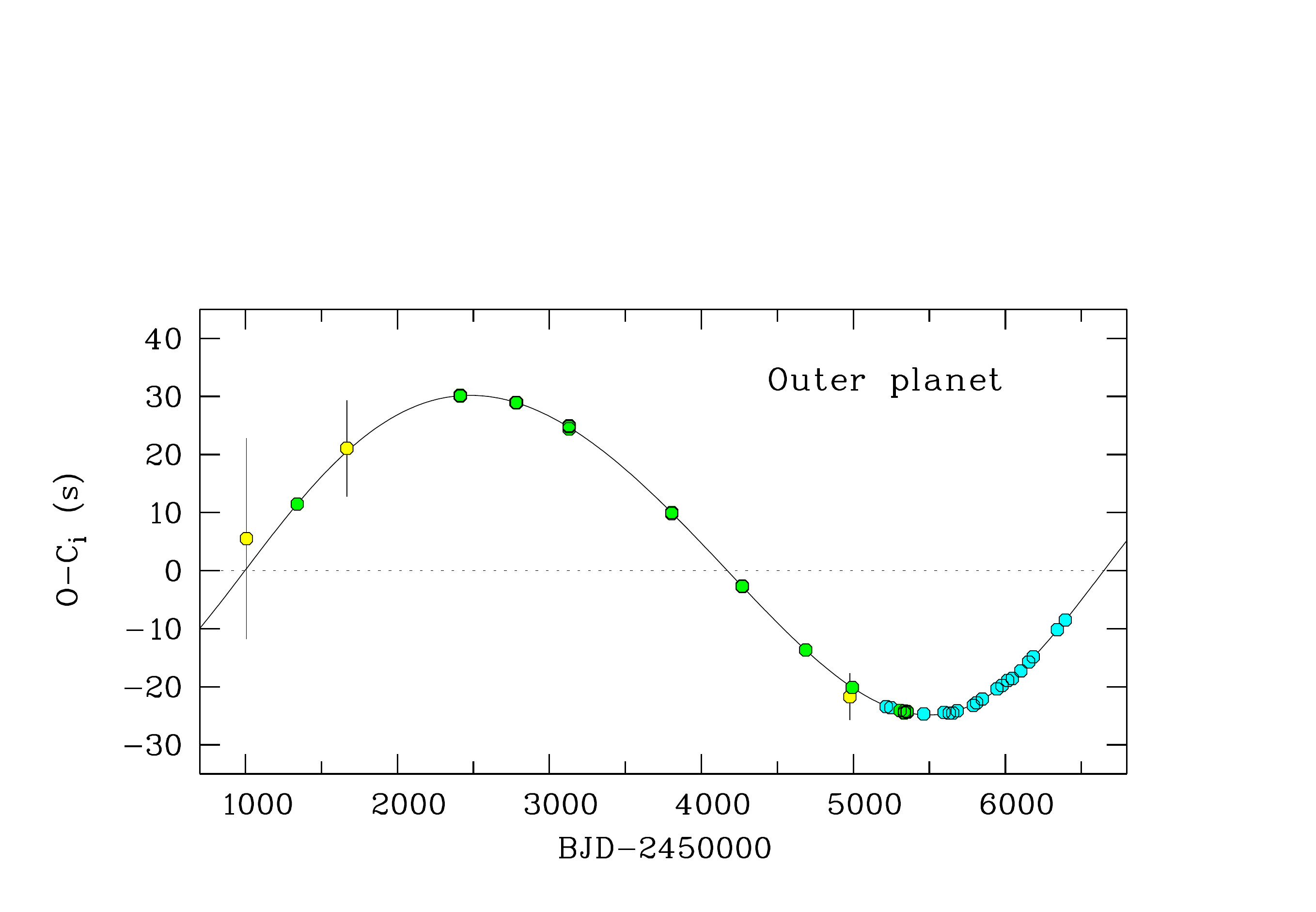}
\includegraphics[bb=168 29 510 689,height=89mm,angle=-90,clip]{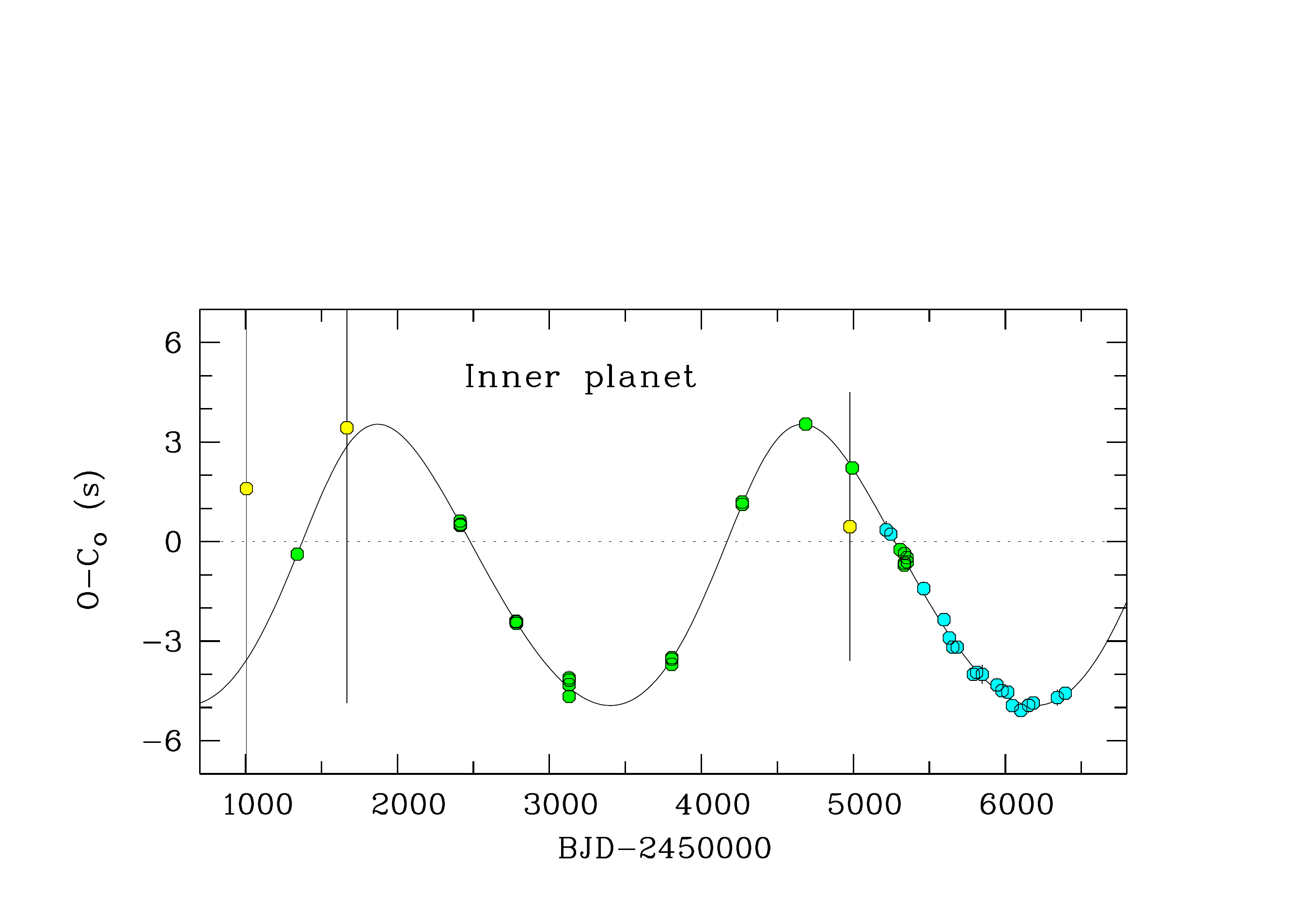}
\caption[chart]{Residuals for the outer planet after subtracting
  the contribution of the inner planet (top) and vice versa
  (bottom). The abscissa divisions are the same for the two panels.
  The residuals from the fit are identical to those of
  Fig.~\ref{fig:oc1}. }
\label{fig:oc2}
\end{figure}

We combined the results of Runs~1 and 2 to investigate the spread of
the fit parameters for the solutions that pass the lifetime criterion,
irrespective of their position in the \eo.\ei\
plane. Figure~\ref{fig:prat} shows the distributions of the orbital
periods \pou\ and \pin\ and the period ratio
$P_\mathrm{o}/P_\mathrm{i}$ for model fits with $\,\chi^2\!<\!109.4$
and $\,\chi^2\!<\!97.9$, corresponding to the selections in the center
and righthand panels of Fig.~\ref{fig:gray}, respectively. For the
more lenient \chisq\ limit, 169 of 173 solutions yield nearly
identical periods and a period ratio of 2.022 with a standard
deviation $\sigma\!=\!0.025$. The remaining four solutions
interestingly have a period ratio of 2.512 with
$\sigma\!=\!0.026$. The requirement of long-term stability implies
that resonant solutions are selected, preferably the 2\,:\,1 case, but
also 5\,:\,2 for a few fits. Other period ratios do not occur and no
nonresonant secularly stable model was found in the entire parameter
space. The stability island, together with all long-lived outliers
(dark red) in the center panel of Fig.~\ref{fig:gray}, represents the
2\,:\,1 case, while the four 5\,:\,2 solutions lie in the light red
region around $e_\mathrm{o},e_\mathrm{o}\!=\!0.12,0.14$ on a \chisq\
ridge. The most longlived of the four has $\tau\!=\!\ten{6.7}{6}$\,yr
with $\,\chi^2\!=\!106.0$ and the best-fitting
$\tau\!=\!\ten{1.1}{6}$\,yr with $\chi^2\!=\!105.6$, clearly inferior
to the island solutions. We exclude the 5\,:\,2 resonant solution at
the 99.3\% confidence level. Truly long-term stable solutions that
provide good fits to the data exist only in the 2\,:\,1 mean-motion
resonance explored in more detail in the next section.

\begin{table}[t]
\begin{flushleft}
  \caption{Observed and derived parameters of the Keplerian two-planet
    LTT model for \nnser.}
\begin{tabular}{l@{\hspace{1mm}}c@{\hspace{4mm}}r}
  \hline \hline\\ [-1ex]
  Parameter & Island solutions\tablefootmark{1} & Best fit\tablefootmark{2}\\ [1ex] 
  \hline \\ 
  \multicolumn{3}{l}{\emph{(a) Observed parameters: }}\\[0.5ex]
  Period $P_\mathrm{o}$~~(yr)   & $15.482\pm 0.027$  & 15.473  \\ 
  Period $P_\mathrm{i}$~~(yr)   & \hspace{1.4mm}$7.647\pm0.058$   & \hspace{0mm} 7.653  \\ 
  Period ratio                & \hspace{1.4mm}$2.025\pm0.018$ &  \hspace{0mm}2.022  \\ 
  Eccentricity $e_\mathrm{o}$   &  \hspace{1.4mm}$0.142\pm 0.011$  &  \hspace{0mm}0.144  \\ 
  Eccentricity $e_\mathrm{i}$   &  \hspace{1.4mm}$0.223\pm0.019$ &  \hspace{0mm}0.222  \\ 
  LTT amplitude $K_\mathrm{bin,o}$~(s)  & $27.65\pm0.12$  &  \hspace{0mm}27.68  \\ 
  LTT amplitude $K_\mathrm{bin,i}$~~(s)   & \hspace{1.4mm}$4.32\pm0.22$ &  4.28 \\ 
  Argument of periapse $\omega_\mathrm{bin,o}$~~($^\circ$) &\hspace{-1.0mm}$317.0\pm\,6.4$ & \hspace{0.0mm}318.4 \\
  Argument of periapse $\omega_\mathrm{bin,i}$~~($^\circ$) & \hspace{0.5mm}$48.0\pm\,2.6$  & \hspace{0.5mm}47.5  \\
  $\chi^2$ for 109 degrees of freedom                 & $<\!97.9$ & 95.56 \\
  Periapse passage $T_\mathrm{o}$~~(BJD)    & \multicolumn{2}{c}{$2\,456\,145\pm80$}\\
  Periapse passage $T_\mathrm{i}$~~(BJD)    & \multicolumn{2}{c}{$2\,454\,406\pm20$}\\[1ex]
  \multicolumn{3}{l}{\emph{(b) Derived parameters: }}\\[0.5ex]
  Outer planet, semi-major axis $a_\mathrm{o}$~(AU) &  \hspace{1.4mm}$5.389\pm0.045$ & 5.387   \\ 
  Inner planet, semi-major axis $a_\mathrm{i}$~(AU) &  \hspace{1.4mm}$3.358\pm0.033$ & 3.360    \\ 
  Outer planet, mass $M_\mathrm{o}$\,sin\,$i_\mathrm{o}$~~(\mjup) & \hspace{1.4mm}$6.96\pm0.12$ & 6.97 \\
  Inner planet, mass $M_\mathrm{i}$\,sin\,$i_\mathrm{i}$~~(\mjup) & \hspace{1.4mm}$1.74\pm0.09$ & 1.73 \\ [1ex]
  \hline
\end{tabular}
\tablefoot{
\tablefoottext{1}{For the island of stability in the righthand panel of Fig.~\ref{fig:gray} with
    $\,\chi^2\!<\!97.9$ and $\tau\!>\!10^6$\,yr, including the
    1-$\sigma$ errors.}
\tablefoottext{2}{For the best fit with
    $\,\chi^2\!=\!95.56$ for 109 degrees of freedom and $\tau\!>\!10^8$\,yr.}
}
\label{tab:nnser}
\end{flushleft}
\end{table}

Reducing the \chisq\ limit to the 1-$\sigma$ level of 97.9 leaves us
with 56 solutions with $\tau\!>\!10^6$\,yr, 44 from Run-1 and 12 from
Run-2 (Fig.~\ref{fig:gray}, righthand panel, and Fig.~\ref{fig:prat},
green histograms). In the multi-dimensional parameter space, these
solutions lie close together and all parameters have narrow
quasi-Gaussian distributions with well-defined mean values, as shown
for the periods and the period ratio in Fig.~\ref{fig:prat}.  Some of
the parameters are strongly correlated. We quote in
Table~\ref{tab:nnser} the mean values and the standard deviations of
the respective parameters with all other parameters free. We searched
for the best fit within the island of stability in Run~3, a subset of
which contains 82 models with $\tau\!>\!10^8$\,yr and $\chi^2$ between
95.56 and 95.60 for 109 degrees of freedom. The parameters of the
best-fitting model are listed separately in Table~\ref{tab:nnser}.
In the bottom part of the table, we list the astrocentric semi-major
axes and planetary masses derived from the observed periods and
LTT-amplitudes, assuming coplanar edge-on orbits.  With masses
sin\,$i_\mathrm{o}\,M_\mathrm{o}\!=\!7.0$\,\mjup\ and
sin\,$i_\mathrm{i}\,M_\mathrm{i}\!=\!1.7$\,\mjup, the two companions
to \nnser\ qualify as giant planets for a wide range of inclinations.

Figures~\ref{fig:oc1} and~\ref{fig:oc2} show the data along with the 
best-fit model of Table~\ref{tab:nnser}.  The ordinate \oclin\ in the
top panel of Fig.~\ref{fig:oc1} is the deviation of the observed
mid-eclipse times from the underlying linear ephemeris of the binary
\begin{equation}
T_\mathrm{ecl}\!=\!\mathrm{BJD(TDB)}\,2447344.524368(7) + 0.13008014203(3)\,E,~~~
\label{eq:ephem}
\end{equation}
which combines the fit parameters $T_\mathrm{bin}$ and
$P_\mathrm{bin}$. The 1-$\sigma$ errors quoted in parentheses reflect
the width of the stability island. The residuals \ocell\ of the 76
individual MONET mid-eclipse times from the fit with two elliptical
orbits are listed in the last column of Table~\ref{tab:monet}. They
have an rms value of 0.34\,s. For clarity in the presentation, we show
the MONET data in Figs.~\ref{fig:oc1} and~\ref{fig:oc2} only in the
form of the weighted mean values for the 19 groups of mid-eclipse
times introduced in Section~\ref{sec:obs} (cyan-blue filled
circles). Their rms value is 0.14\,s. If a third periodicity hides in
the residuals, its amplitude does not exceed
0.25\,s. Figure~\ref{fig:oc2} shows the contributions \oci\ of the
outer and \oco\ of the inner planet to the LTT signal. The data points
in these graphs are obtained by subtracting the LTT contribution of
the mutually other planet in addition to the linear term from the
observed mid-eclipse times. It is remarkable how well the two-planet
model fits the data, since the observations now cover nearly a
complete orbit of the outer planet and two orbits of the inner one.

\subsection{Temporal evolution of the planetary system of \nnser}
\label{sec:model}

In this section, we explore the temporal evolution of the secularly
stable two-planet models that start with orbital elements defined by
the Keplerian fits to the data. We find that all long-lived models of
Fig.~\ref{fig:gray} (central and righthand panels) adhere to the
2\,:\,1 mean-motion resonance. This holds for the models in the
stability island, but also for the solutions that correspond to the
skerries surrounding it. Their isolated character probably results
from the increasing difficulty of finding a set of parameters that
secures resonance, as the start parameters deviate from their optimal
values. In all these solutions, the mean-motion resonant variable
$\Theta_1\!=\!\lambda_\mathrm{i}-2\,\lambda_\mathrm{o}+\omega_\mathrm{i}$,
a function of the planet longitudes $\lambda_\mathrm{i}$ and
$\lambda_\mathrm{o}$, librates about zero and the secular variable
$\Delta \omega\!=\!\omega_\mathrm{i}-\omega_\mathrm{o}$
circulates. This agrees with the expected behavior of a two-planet
system with a mass ratio $M_\mathrm{o}/M_\mathrm{i}\!>\!1$
\citep{michtchenkoetal08}.

% Fig. 7
\begin{figure*}[t]
\includegraphics[bb=184 67 466 722,height=90mm,angle=-90,clip]{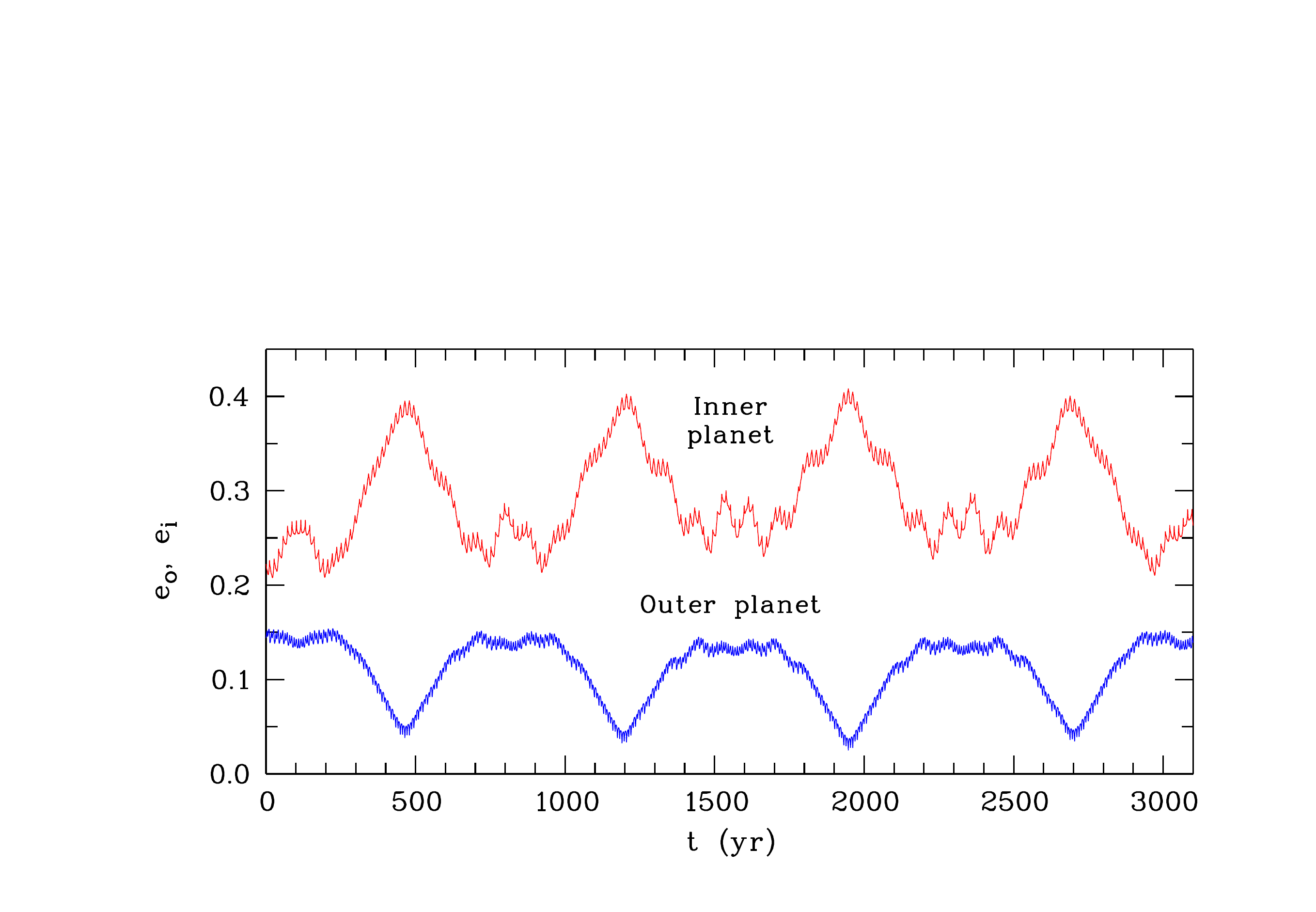} \hfill
\includegraphics[bb=184 67 466 722,height=90mm,angle=-90,clip]{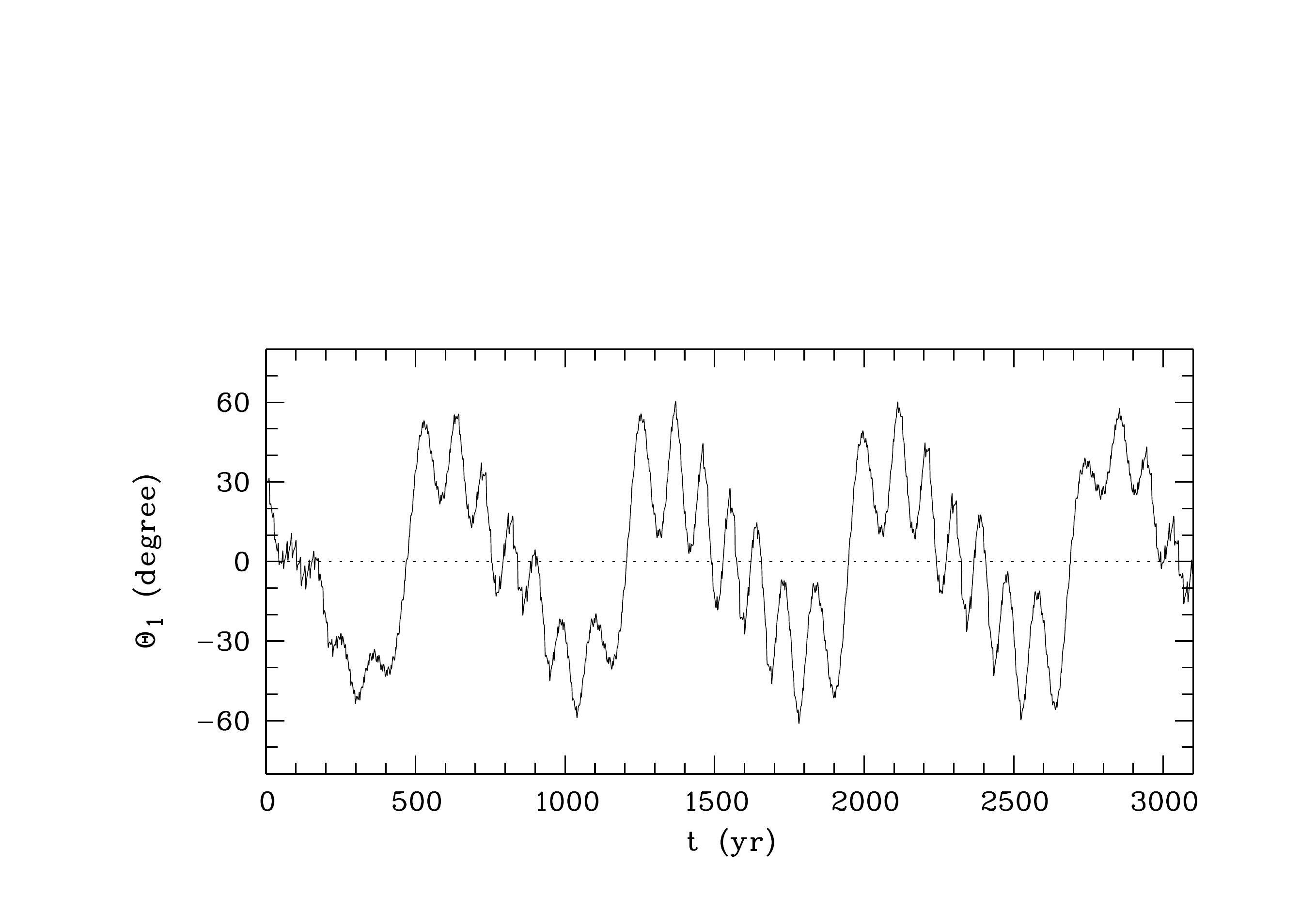}
\includegraphics[bb=204 67 509 722,height=90mm,angle=-90,clip]{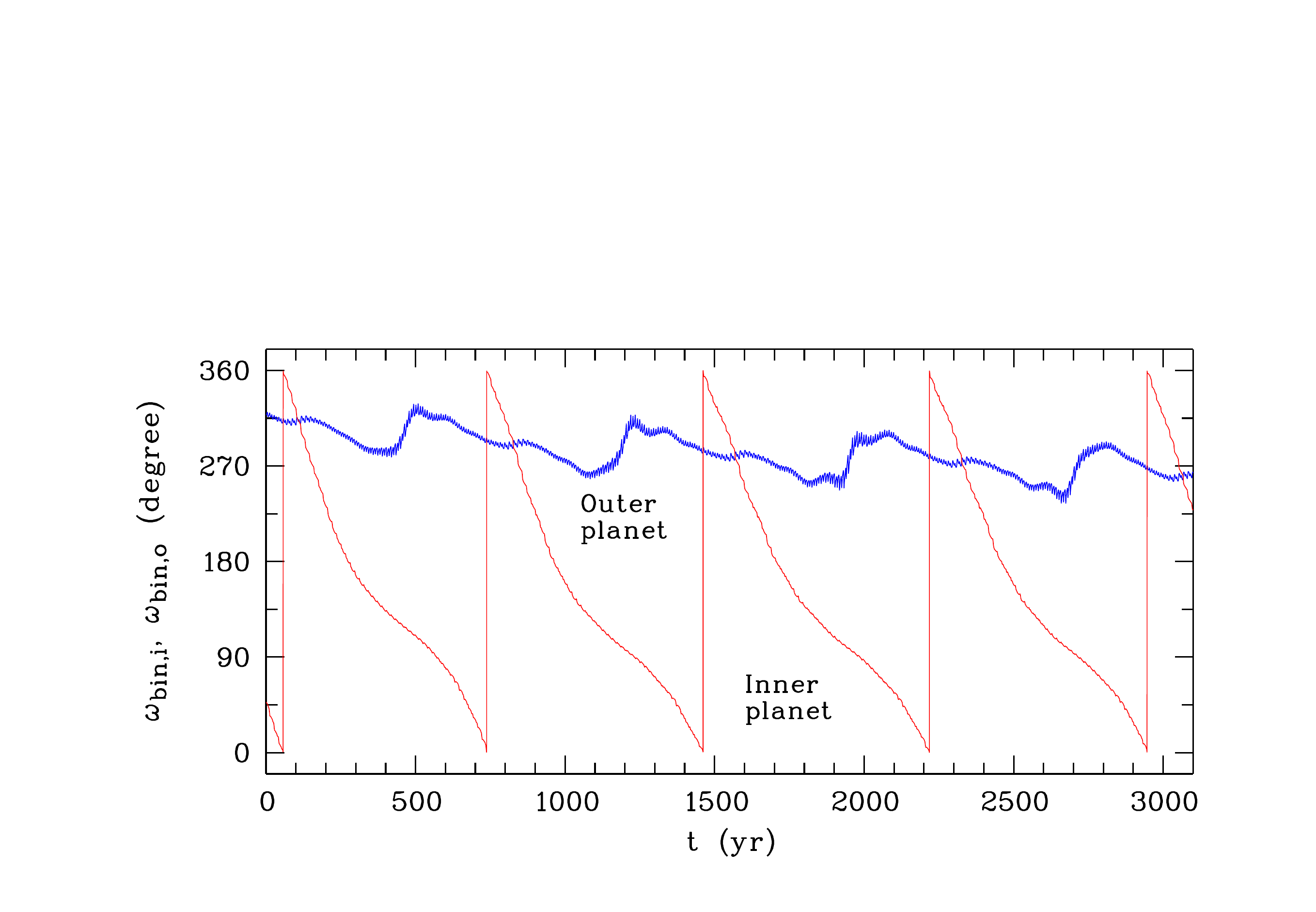} \hfill
\includegraphics[bb=204 67 509 722,height=90mm,angle=-90,clip]{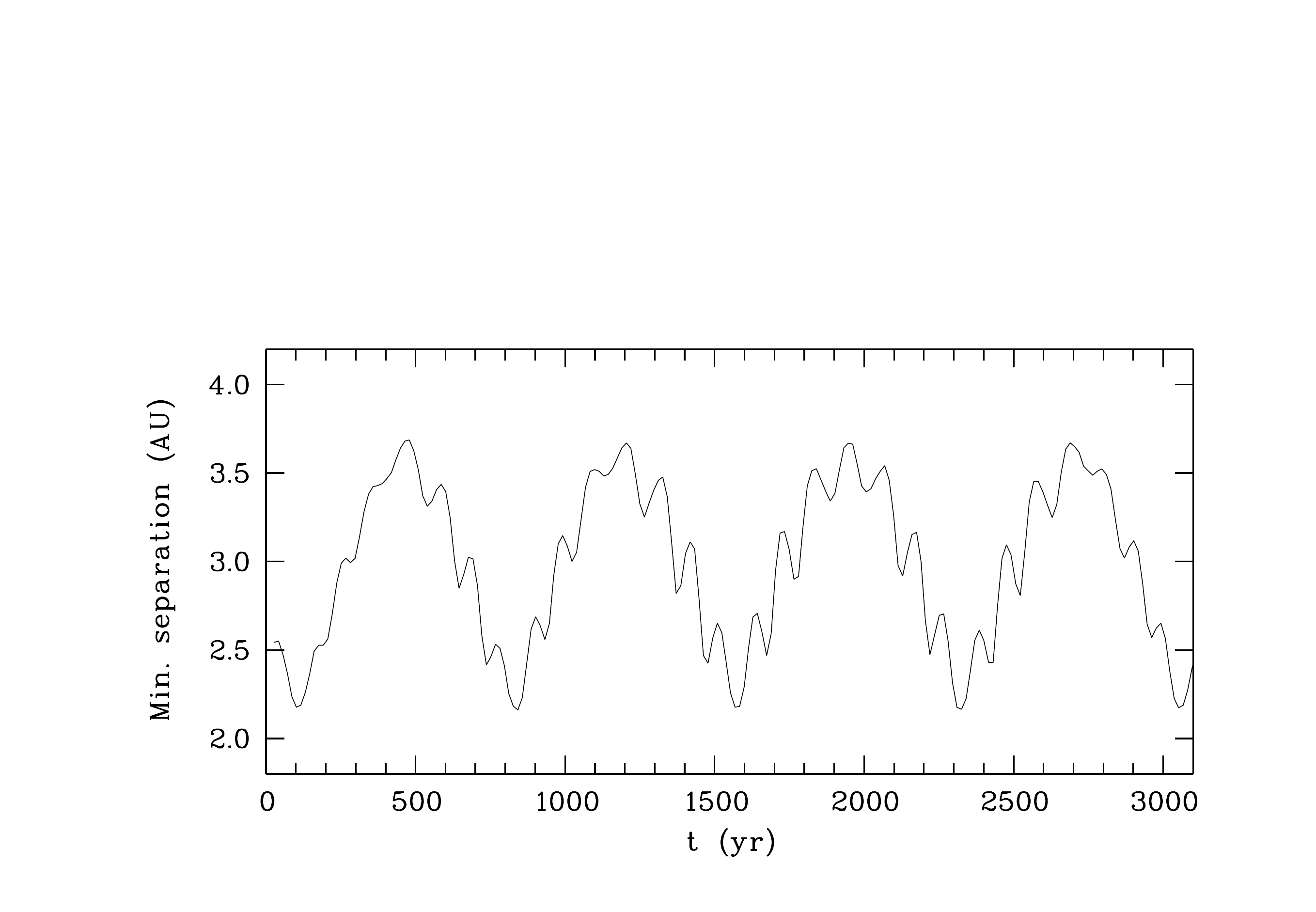}
\caption[chart]{Temporal evolution of selected osculating orbital
  elements of a model, starting from the best-fit Keplerian parameters
  of Table~\ref{tab:nnser}. The first 3050 yr of the greater than
  $10^8$\,yr lifetime are displayed. The left-hand panels show the
  eccentricities and arguments of periapse for the motion of the
  center of mass of the binary, the top righthand panel shows the
  resonant angle $\Theta_1$, and the bottom righthand panel the
  minimum separation between the two planets for successive
  orbits of the inner planet.}
\label{fig:model}
\end{figure*}

\begin{figure}[t]
\includegraphics[bb=51 90 509 580,height=80mm,angle=-90,clip]{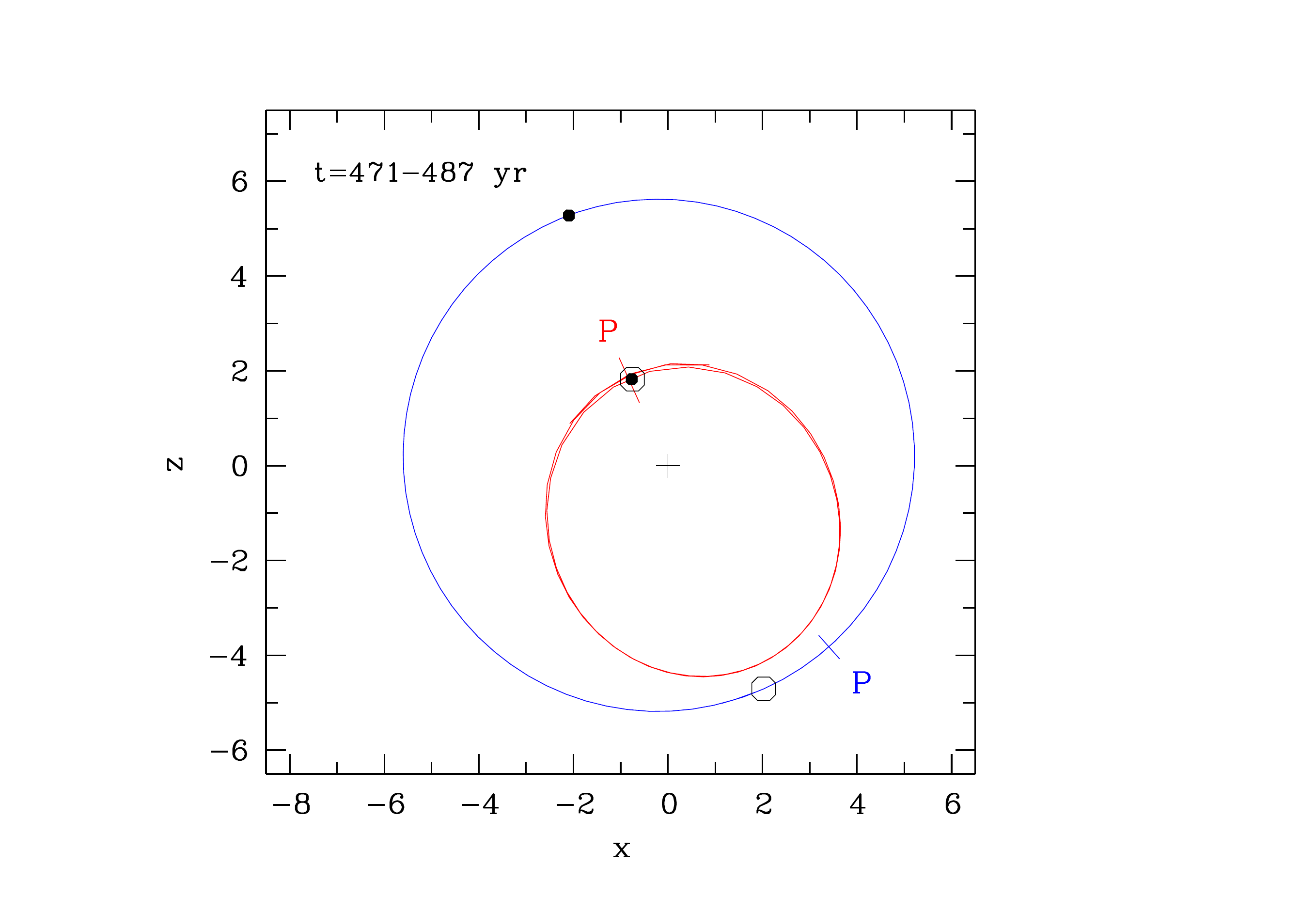}
\caption[chart]{Astrocentric orbits for $t\!=\!471-487$\,yr after the
  begin of the calculation. The locations of the periapses are marked
  `P', the solid dots indicate conjugation, and the open circles
  opposition. Orbital motion is counter-clockwise.}
\label{fig:orbits}
\end{figure}

Figure~\ref{fig:model} shows the evolution of selected parameters for
the best-fitting model of Table~\ref{tab:nnser} over the first
3050\,yr of the lifetime, which exceeds $10^8$\,yr. The two principal
periods of the system \citep{reinpapaloizou09} are 105\,yr and
736\,yr. Both are prominent in the librations of $\Theta_1$. In the
shorter period, low-amplitude anti-phased oscillations of the
semi-major axes occur (not shown). In the longer period, $\Delta
\omega$ circulates, anti-phased oscillations of the eccentricities
take place, and the minimum separation between the two planets
reoccurs. The protection mechanism due to the 2\,:\,1 resonance always keeps
the separation above 2.15\,AU, effectively limiting the mutual
gravitational interaction (bottom right panel). As an illustration, we
show the orbits at the time of peak eccentricity $e_\mathrm{i}$ in
Fig.~\ref{fig:orbits}.  The system is locked deep in the 2\,:\,1
mean-motion resonance, as demonstrated by the near-zero mean values of
$\Theta_1$, averaged over successive 736-yr intervals. For the first
20,000\,yr of the evolution, the 26 values of $\langle\Theta_1\rangle$
can be represented by an underlying linear and a superposed sinusoidal
variation with a period of 3450\,yr and an amplitude of
$0.86^\circ$. The linear component has a fitted slope of $\langle\dot
\Theta_1\rangle\!=\!\ten{(-0.2\pm1.7)}{-6}$\,deg\,yr$^{-1}$, entirely
consistent with zero. All long-lived solutions that start from fits in
the stability island behave similarly to the best-fit model, in the sense
that all of them have $\Theta_1$ librating and $\Delta \omega$
circulating. This also holds for the longlived outliers in the central
panel of Fig.~\ref{fig:gray} that have $\tau\!>\!10^8$\,yr. They
differ in the secular period, which ranges from 330 to 1700\,yr.

% Fig. 8

All models considered thus far involved prograde coplanar edge-on
orbits. We calculated a few models with different inclinations
$i_\mathrm{i}$ and $i_\mathrm{o}$ of the inner and outer planet as a
first step toward a more comprehensive study of \nnser's planetary
system. A common tilt in the planetary orbits, which enhances the
planetary masses is limited to $25^\circ$, beyond which instability
increases rapidly. Similarly, the calculations limited the mutual tilt
of the two orbits with respect to each other to $5^\circ$.

\section{Summary and discussion}

We have presented 69 new mid-eclipse times of \nnser, which cover the
minimum and subsequent recovery of the \oclin\ residuals expected from
our previous model in Paper~I. The data allowed us to derive a
significantly improved two-planet model for \nnser\ based on the
interpretation of the observed eclipse-time variations in terms of the
LTT effect. Combined with extensive stability calculations, we find
that the only model that simultaneously fits the data and is secularly
stable involves two planets locked in the 2\,:\,1 mean-motion
resonance.  We did not find any good fits that are nonresonant and
stable. Apart from differences in the periods and mass ratio, the
\nnser\ system bears resemblance to the HD\,128311 planetary system
\citep{sandorkley06,reinpapaloizou09}. The best fit implies
eccentricities $e_\mathrm{o}\!=\!0.14$ and $e_\mathrm{i}\!=\!0.22$ for
the outer and inner planets, respectively, and a mass ratio
$M_\mathrm{o}/M_\mathrm{i}\,=\,4$ for coplanar orbits. As expected
theoretically for a system with a more massive outer planet
\citep{michtchenkoetal08}, the temporal evolution of the model that
starts from our best fit involves a libration of the resonant variable
$\Theta_1$ and a circulation of $\Delta \omega$. Preliminary stability
calculations for tilted orbits suggest that deviations from
coplanarity with the binary orbit cannot be large.

Eclipse-time variations in PCEB seem to be ubiquitous \citep[e.g.][and
references therein]{beuermannetal12,parsonsetal10b,qianetal12} and the
plausibility of explaining them by the LTT effect has increased by the
definitive discovery of planets orbiting non-evolved close binaries
with KEPLER \citep[e.g.][]{doyleetal11,oroszetal12,welshetal12}. That
the orbital periods in the two types of systems differ by one to two
orders of magnitude is a selection effect, because the LTT effect
increases with the orbital period, while the radial velocities and
transit probabilities decrease. Contrary to radial velocities
measured by line shifts, however, the interpretation of the
eclipse-time variations in terms of a displacement of the binary along
the line of sight is not unique, since eclipse-time variations can
also be produced by mechanisms internal to the binary
\citep[e.g.][]{applegate92}. Most authors, however, consider this
action as inadequate for explaining the magnitude of the observed variations
\citep[e.g.][]{brinkworthetal06,chen09,watsonmarsh10}. Doubts about
the LTT interpretation nevertheless remain, raised by such problematic
cases as \huaqr\ \citep{gozdziewskietal12,horneretal11,hinseetal12}
and QS~Vir \citep{parsonsetal10b}. Resolving these cases remains an
important task. Presently, \nnser\ represents the best documented case
in favor of the LTT hypothesis with an agreement between data and
model at the 100\,ms level, and it is also the prime contender for an
observational proof of the required strict periodicity of the signal,
given enough time.

The evolutionary history of planetary systems orbiting PCEB may be
complex. Planets either existed before the CE and had to pass through
the envelope or they were formed in it. In the former case, the orbit
of any pre-existing planet is severely affected by the loss of
typically more than one half of the mass of the central object. In the
latter case, formation depends critically on the conditions in the
expanding envelope. In both scenarios, migration may drive a planet
pair into resonance, but predicting the properties of the emerging
post-CE system is a difficult task. Establishing the structure of systems
like \nnser\ may provide the observational basis for such a program.

\begin{acknowledgements}
  We thank the referee for the careful reading of the paper and the
  constructive and helpful comments, which served to significantly
  improve the presentation. This work is based in part on data
  obtained with the MOnitoring NEtwork of Telescopes (MONET), funded
  by the Alfried Krupp von Bohlen und Halbach Foundation, Essen, and
  operated by the Georg-August-Universit\"at G\"ottingen, the McDonald
  Observatory of the University of Texas at Austin, and the South
  African Astronomical Observatory. We thank George Miller (UT
  Austin), Bernhard Starck (Gymnasium Frankenberg), Ronald Sch\"unecke
  (Evangelisches Gymnasium Lippstadt), Paul Breitenstein (Gymnasium
  M\"unster) and Alexander Schmelev (Institut f\"ur Astrophysik
  G\"ottingen) for taking a total of ten of the eclipses listed in
  Table~1, some of them as part of the MONET \emph{PlanetFinders}
  program that involves high school teachers and students.
\end{acknowledgements}

\bibliographystyle{aa}

\end{document}